\documentclass[twocolumn,prb,amsmath,amssymb,floatfix,groupedaddress]{revtex4-2} 
\usepackage{graphicx} 
\usepackage{bm}
\usepackage{bbm}
\usepackage{color}
\usepackage{amsmath}
\usepackage{amssymb}
\usepackage{color}
\usepackage{footnote}
\usepackage{comment}
\usepackage{hyperref}
\usepackage{subfigure}
\usepackage{appendix}
\usepackage[applemac]{inputenc}

\hypersetup{ colorlinks, citecolor=blue, linkcolor=blue, urlcolor=blue}

\begin{document}
	\title{Spin-polarized Andreev molecules and anomalous nonlocal Josephson effects in altermagnetic junctions}
	
	\author{Sayan Mondal}
	\author{Jorge Cayao}
	\thanks{Contact author: jorge.cayao@physics.uu.se}
	\affiliation{Department of Physics and Astronomy, Uppsala University, Box 516, S-751 20 Uppsala, Sweden}

	\date{\today} 
	\begin{abstract}
		Altermagnetism has emerged as a promising ingredient for realizing nontrivial Josephson phases but so far explored   in single Josephson junctions. In this work, we consider the coherent coupling of two Josephson junctions with spin-singlet $s$-wave superconductivity and demonstrate that $d$-wave  altermagnetism gives rise to spin-polarized Andreev molecules due to the hybridization of Andreev bound states of each junction when the coupling is weak. Interestingly, these spin-polarized Andreev molecules induce anomalous nonlocal Josephson effect, where the current flow across one Josephson junction due to phase changes across the other  junction develops  $0-\pi$ and $\phi_{0}$ transitions originating from altermagnetism.
		Furthermore, the nonlocal Josephson current carried by spin-polarized Andreev molecules   exhibits nonreciprocal critical currents, enabling a nonlocal Josephson diode effect whose polarity is tunable by the   altermagnetic strength and right phase. Our findings put forward altermagnetism as a promising arena for designing nonlocal spin Josephson phenomena. 
		
	\end{abstract}
	\maketitle
	
	\section{Introduction}
	The advent of altermagnetism has provided an unexplored front in the search for nontrivial
	superconducting phenomena that are promising for quantum applications \cite{FukayaJPCM2025,liu2025review}. This is  because altermagnets (AMs) possess anisotropic spin-polarized Fermi surfaces and zero net magnetization \cite{noda2016momentum,NakaNatCommun2019,Hayami19,Ahn2019,Yuanprb20,LiborSAv,NakaPRB2020,Yuanprm21,LiborPRX22,MazinPRX22,landscape22, ma2021multifunctional}, originated from a  collinear-compensated magnetic ordering in real space   that preserves inversion symmetry but breaks time-reversal symmetry (TRS). While these properties enabled the prediction of a number of unusual superconducting phases \cite{FukayaJPCM2025,liu2025review,mazin2025notes,song2025altermagnets}, their role in   Josephson systems  is of particular relevance because Josephson junctions (JJs) form the basis for superconducting electronics and quantum computing \cite{devoret2004superconducting,clarke2008SC,Brecht2016,krantz2019quantum,martinis2020quantum,aguado2020perspective,benito2020hybrid,aguado2020majorana,Cayao2020odd,PRXQuantum.2.040204,siddiqi2021engineering,bargerbos2022singlet,doi:10.1146/annurev-conmatphys-031119-050605,pita2023direct,tanakaReview2024}. In this regard, AMs in single JJs have already proven crucial for inducing spin-polarized Andreev bound states (ABSs) \cite{PhysRevLett.131.076003,PhysRevB.108.075425,PhysRevB.111.064502,vosoughinia2025altermon}, anomalous supercurrents \cite{zhang2024finite,PhysRevLett.131.076003, yang2026topological,PhysRevLett.133.226002,PhysRevB.109.024517,PhysRevB.111.064502,PhysRevB.111.165406,PhysRevB.111.184515,mj4b-2fnr,prnx-47mk}, and nonreciprocal critical currents \cite{PhysRevB.110.014518,Jiang_2025,yqsg-xdg8,debnath2025,sharma2026pUM,esin2026JDE,shaffer2025SDE}.  These advances   show the potential of altermagnetism in single JJs, and naturally make  us wonder about the impact of altermagnetism on  coherently coupled   JJs.

	The coherent coupling of JJs  has already inspired several theoretical  \cite{Pillet_2019,10.21468/SciPostPhysCore.2.2.009,PhysRevB.97.035443,PhysRevB.102.245435,PhysRevB.104.075402,PhysRevB.108.174502,PhysRevB.109.205406,PhysRevResearch.5.033199,PRXQuantum.5.020340,PhysRevB.109.245133,PhysRevB.110.235426,10.21468/SciPostPhys.17.2.037,kotetes2024nonRecifourpi,PhysRevB.111.024506,cayao2026nonlocal} and experimental \cite{strambini2016omega,Su_2017,draelos2019supercurrent,PhysRevX.10.031051,graziano2022selective,Matsuo_2022,junger2023intermediate,matsuo2023phase,matsuo2023phase2,Haxell_2023,PRXQuantum.5.020301,zhu2025josephson} efforts. Central to these works is the formation of Andreev molecules due to the hybridization of ABSs when coupling two JJs  \cite{Pillet_2019}, a  phenomenon tunable  by the superconducting phase differences across each JJ. Andreev molecules form  similarly   to how natural atoms bond to form chemical molecules  but   based on the Andreev and Josephson physics. A unique property of the Andreev molecular regime is that the supercurrent flow  across one JJ can be manipulated by the phase difference across the other JJ, enabling   the so-called nonlocal Josephson effect (JE) \cite{Pillet_2019}. At present, there already exists experimental evidence of Andreev molecules and of the nonlocal JE, thus supporting coherently coupled JJs  for creating fundamental states and intriguing material properties. Despite this progress, however, the research on Andreev molecules and nonlocal JE has mainly focused on superconductor-semiconductor systems, leaving seldom studied the role of the recently discovered altermagnetism.

	In this work, we consider three  coupled spin-singlet $s$-wave superconductors with $d$-wave altermagnetism  and demonstrate the formation of spin-polarized Andreev molecules
	when spin-polarized ABSs of each JJ hybridize across the middle superconductor. 
	Notably, we find that the spin-polarized Andreev molecules enable anomalous phase-biased nonlocal Josephson effects, whereby a phase difference across one junction induces a supercurrent through the other.  Depending on the  altermagnetic crystal symmetries and strength of altermagnetism, the nonlocal Josephson current develops tunable transitions between $0$-, $\pi$-, and $\phi_0$-junction behaviors. 
	We also show that the nonlocal supercurrents carried by spin-polarized Andreev molecules develop a nonreciprocal behavior, thereby signalling the emergence of a nonlocal Josephson diode effect. We find that the efficiency and polarity of this diode effect exhibit a strong tunability by the type and strength of altermagnetism. 
The core feature unifying these diverse phenomena is the emergent, spin-polarized nature of the Andreev molecules. Because $d$-wave altermagnetism spin splits the Fermi surface in a momentum-dependent way while maintaining zero net magnetization, the hybridized Andreev bound states inherit a unique altermagnetic spin polarization.  This resulting spin polarization acts as the singular catalyst for the anomalous nonlocal Josephson response, the $0$-, $\pi$-, and $\phi_{0}$-junction transitions, and the nonlocal Josephson diode effect alike.
Our work establishes the interplay between altermagnetism and coherently coupled JJs as an outstanding ground for designing nonlocal   Josephson phases with spintronic functionalities.
	
The remainder of this work is organized as follows. In Sec.\,\ref{sec:Coupled_Josephson_Junctions_AM}, we present the model of coherently coupled Josephson junctions with $d$-wave altermagnetism. In Sec.\,\ref{sec:spin-polarized_Andreev_molecules}, we demonstrate the emergence of spin-polarized Andreev molecules, while in Secs.\,\ref{sec:Anomalous_nonlocal_Josephson_currents} and \ref{sec:Nonlocal_JDE} we discuss their impact on the nonlocal Josephson effect and the nonlocal Josephson diode effect, respectively. In Sec.\,\ref{sec:conclusions}, we present our conclusions.

	\section{Coupled Josephson junctions with $d$-wave altermagnetism}
	\label{sec:Coupled_Josephson_Junctions_AM}
	
\begin{figure*}[!t]
\centering
\includegraphics[width=\linewidth]{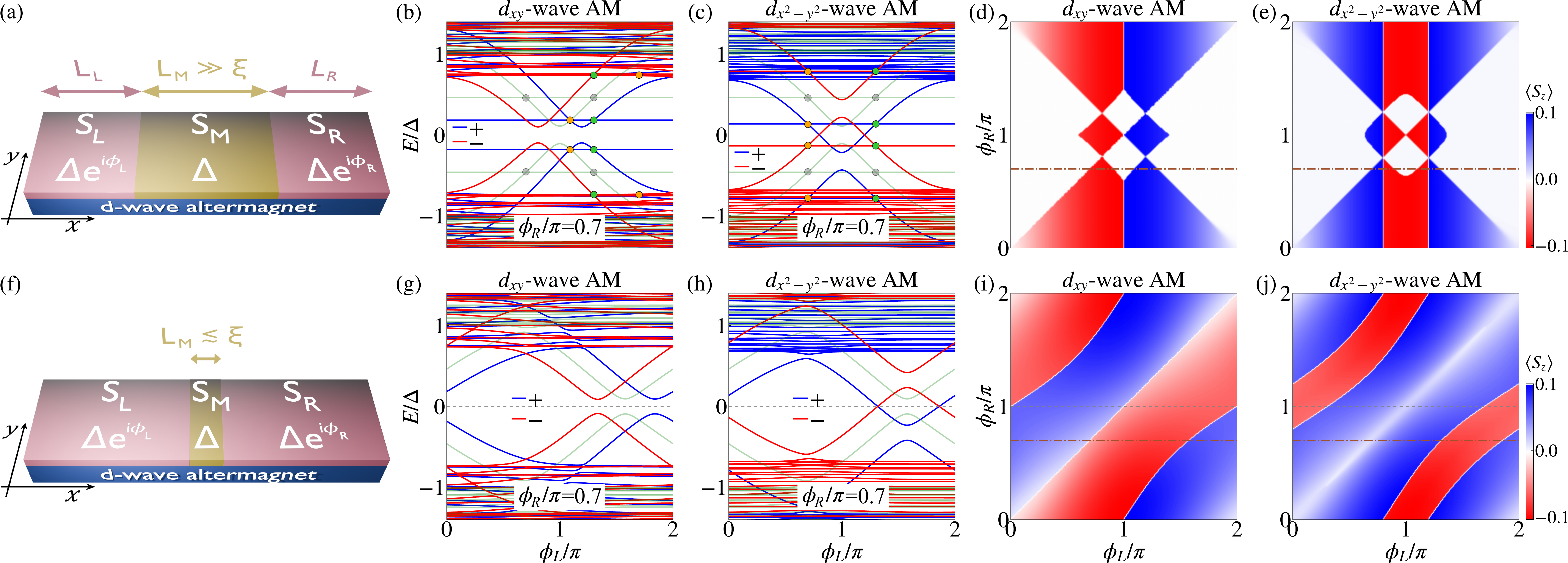}
\caption{(a,f) Sketch of the studied Josephson setup, where three spin-singlet $s$-wave superconductors denoted by $S_{\alpha}$ ($\alpha={\rm L,R}$ pink and $\alpha={\rm M}$ gold)  are in contact with a $d$-wave altermagnet (blue);  the superconductors have lengths $L_{\alpha}$ and  pair potentials $\Delta_{\rm \alpha}=\Delta {\rm e}^{i\phi_{\alpha}}$, with $\phi_{\rm M}=0$.   (a) and (f) correspond to  $S_{\rm M}$ with a long ($L_{\rm M}=100a\gg\xi$) and  short ($L_{\rm M}=2a\lesssim\xi$) length, where $\xi$ is the superconducting coherence length   and $a$ is the lattice spacing. (b,c) Low-energy spectrum as a function of $\phi_{\rm L}$ at $\phi_{\rm R}=0.7\pi$ and  $k_{y}=0.1\pi$ for  $L_{\rm M}\gg\xi$, while (g,h) for $L_{\rm M}\lesssim\xi$. The blue (red) color in   (b,c,g,h) indicates that such energy levels belong to the Nambu sector   formed by spin-$\uparrow$ electrons (holes) with spin-$\downarrow$ holes (electrons),   denoted by `$+$' (`$-$').  The light green curves are the energy levels without altermagnetism. The filled yellow and green circles mark the positions where dispersionless levels of the right JJ cross with the dispersive levels of the left JJ   in presence of the AM, while the filled gray circles mark the same but in absence of the AM. (d,e) Spin projection along $z$ in the left JJ at  $k_{y}=0.1\pi$ for  $L_{\rm M}\gg\xi$, while (i,j) for $L_{\rm M}\lesssim\xi$.
Parameters: $L_{\rm L} = L_{\rm R} = 100a$,  $\mu = 1.5t$, $\Delta = 0.2t$, $J_1 = 0.1t$, $J_2 = 0.1t$, and $\phi_{\rm M} = 0$.}
\label{Fig1} 
\end{figure*}
	
	We consider the coherent coupling of two JJs with $d$-wave  altermagnetism, where  two spin-singlet $s$-wave superconductors are coupled through a  spin-singlet $s$-wave superconductor [Figs.\,\ref{Fig1}(a,f)]. The coupled JJs are assumed to be finite along   $x$ and periodic along the $y$, and   modelled by 
	\begin{equation}
		\label{eqJJ}
		{\cal H}_{k_y}(\phi_{\rm L}, \phi_{\rm M}, \phi_{\rm R}) = \sum_{\alpha} H_{k_{y}}(\phi_\alpha) + H_{\rm LM} + H_{\rm MR}\,,
	\end{equation}
	where
	\begin{equation}
		\label{eq:Hk}
		\begin{split}
			H_{k_y}(\phi_\alpha) &= \sum_{i} c_{i}^\dagger h_{ii}(k_y) c_{i} + \sum_{\langle i,j \rangle} c_{i}^\dagger v_{ij}(k_y) c_{j} \\
			&+\sum_{i} \Delta_{\alpha}  (c^{\dagger}_{i_\alpha, \uparrow} c^{\dagger}_{i_\alpha, \downarrow} - c^{\dagger}_{i_\alpha, \downarrow} c^{\dagger}_{i_\alpha, \uparrow}) + \mathrm{H.c.}\,,
		\end{split}
	\end{equation}
	describes the left (L), middle (M), and right (R) superconductor ($\alpha={\rm L,M,R}$) with $d$-wave altermagnetism.
	Here, $c_{i} = \left( c_{i, \uparrow}, c_{i, \downarrow} \right)^{\rm T}$, where $c_{i, \uparrow}$ annihilates an electronic state with spin $\sigma=\uparrow, \downarrow$ at $i$. Moreover, $h_{ii}(k_y) = \left[ -\mu +4t -2t\cos k_y \right] \sigma_0 - J_2 \cos k_y \sigma_z$ and $v_{ij}(k_y)= -t \sigma_0 - i J_1 \sin k_y \sigma_z + (J_2/2) \sigma_z$  correspond to on-site  and nearest-neighbor terms,  which are  block diagonal in spin space,  $\sigma_i$ are the spin Pauli matrices, $t$ denotes the hopping strength, $\mu$ is   the chemical potential, while  $J_1$ and $J_2$  represent   the strength of the $d_{xy}$- and $d_{x^2-y^2}$-wave altermagnetism. Furthermore,  $\Delta_{\alpha}=\Delta e^{i\phi_\alpha} $ is the pair   potential in $\alpha$ superconductor; by gauge symmetry \cite{cayao2026nonlocal}, we set $\phi_{\rm M}=0$ and keep  $\phi_{\rm L,R}\neq0$ characterizing the superconducting phase differences across the left and right JJs. 
We treat $\phi_{\rm L}$ and $\phi_{\rm R}$ as independent control parameters, as  in multiterminal Josephson structures \cite{riwar2016multi,Pillet_2019}, where the superconducting phase differences are biased independently by threading magnetic fluxes through superconducting loops attached to the superconductors. We assume an infinitely wide junction geometry with perfect translational invariance along the $y$ axis, rendering $k_{y}$ a conserved quantum number \cite{PhysRevX.7.021032,PhysRevLett.133.226002,yang2026topological}. This approximation remains highly accurate because any phase-biasing circuitry or loop boundaries are localized at the macroscopic edges, far from the mesoscopic transport channel. Furthermore, since the altermagnetic spin splitting occurs at zero external magnetic field \cite{FukayaJPCM2025}, the local Hamiltonian remains strictly uniform along the interface. 
The term $H_{\rm LM(MR)}$ couples the left and middle (middle and right) superconductors, having non-zero values only between the   sites adjacent to the interfaces. Next, we denote ${\cal H}_{k_y}(\phi_{\rm L}, \phi_{\rm R})\equiv{\cal H}_{k_y}(\phi_{\rm L}, \phi_{\rm M}=0, \phi_{\rm R})$.
	
	Before going further, we   discuss the symmetries of the JJ setup in Eq.\,(\ref{eqJJ}). Without the AM, the JJ Hamiltonian preserves global time-reversal symmetry,
	${\cal T}{\cal H}(\phi_{\rm L},\phi_{\rm R}){\cal T}^{-1} = {\cal H}(-\phi_{\rm L},-\phi_{\rm R})$,
	but local TRS can be broken,
	${\cal T}{\cal H}(\phi_{\rm L},\phi_{\rm R}){\cal T}^{-1} \neq {\cal H}(-\phi_{\rm L},\phi_{\rm R})$, where ${\cal T}$ denotes the time-reversal operator. In the presence of an AM,  global TRS is broken. In addition, in the presence and absence of AM, the JJ intrinsically breaks local spatial inversion, ${\cal I}{\cal H}(\phi_{\rm L},\phi_{\rm R}){\cal I}^{-1} \neq {\cal H}(-\phi_{\rm L},\phi_{\rm R})$, with ${\cal I}$ the inversion operator. Thus, breaking local time-reversal and local inversion symmetries is very likely to lead to unusual ABSs and supercurrents; in fact, the simultaneous breaking of these symmetries satisfies the necessary conditions for   the Josephson diode effect \cite{PhysRevLett.99.067004, PhysRevB.92.035428, PhysRevB.93.174502, PhysRevB.103.245302, PhysRevB.106.214524, davydova2022universal, 79tj-c3y4, PhysRevB.109.L081405, PhysRevResearch.6.L022002, PhysRevB.110.014519, zhang2022}, suggesting that our JJ in Eq.\,(\ref{eqJJ}) can host intriguing Josephson physics.

	\section{Spin-polarized Andreev molecules}
	\label{sec:spin-polarized_Andreev_molecules}
	
	We begin by inspecting the role of the superconducting phase differences $\phi_{\rm L,R}$ in the low-energy spectrum of our coupled JJs with altermagnetism   [Eq.\,(\ref{eqJJ})]. For this reason, Figs.\,\ref{Fig1}(b,c) show  the low-energy spectrum as a function of $\phi_{\rm L}$ with $d_{x^{2}-y^{2}}$- and $d_{xy}$-wave altermagnetism; in both cases,  $\phi_{\rm R}=0.7\pi$, $k_{y}=0.1\pi$ for a   middle superconductor longer than the superconducting coherence length $\xi$
	($L_{\rm M}\gg\xi$). To contrast this regime, Figs.\,\ref{Fig1}(g,h) show the low-energy spectrum for a short middle superconductor ($L_{\rm M}\lesssim\xi$).  These short and long middle superconductor  regimes capture the weak and strong couplings between JJs. To complement the analysis in these two regimes, Figs.\,\ref{Fig1}(d,e,i,j)  show the spin polarization   $\langle S_z \rangle$ in the left JJ for the first positive ABS.   In the absence of altermagnetism, the left (right) JJ  host four spin-degenerate ABSs that depend on $\phi_{\rm L,R}$ and $L_{\rm M}$, with a $2\pi$-periodic energy spectrum that is symmetric about $\phi_{\rm L}=\pi$ [see   light green curves in Figs.\,\ref{Fig1}(b,c,g,h)]. For $L_{\rm M}\gg\xi$, the left and right JJs behave  as independent [Figs.\,\ref{Fig1}(b,c)]:    The ABSs of the left JJ exhibit a strong dependence on $\phi_{\rm L}$, while the ABSs of the right JJ remain dispersionless and intersect the ABSs of the left JJ at $\phi_{\rm R} = \phi_{\rm L} $, as indicated by the gray circles in Fig.\,\ref{Fig1}(b). In contrast, for $L_{\rm M}\lesssim\xi$ [Figs.\,\ref{Fig1}(g,h)], the ABSs of both junctions hybridize, leaving only a single pair of hybridized ABSs  that strongly disperses with $\phi_{\rm L,R}$  and is asymmetric with respect to $\phi_{\rm L}=\pi$. The dependence of the hybridized ABSs on $L_{\rm M}$ resembles an interatomic potential describing molecular formation, giving rise to the so-called Andreev molecules \cite{Pillet_2019}; see Fig.\,\ref{FigE2} in App.\,\ref{app:Molecular behavior of the hybridized Andreev states} for the molecular energy dependence on $L_{\rm M}$ and Fig.\,\ref{FigE1} in App.\,\ref{app:localization of Andreev molecules} for the real-space wave function hybridization.

	With altermagnetism, the low-energy spectrum develops interesting changes in short and long middle superconductors [see Figs.\,\ref{Fig1}(b,c,g,h)]. First, a finite amount of altermagnetism makes ABSs to split in spin,   giving rise to spin-polarized ABSs with a behavior that   strongly depends on the ratio between $L_{\rm M}$ and $\xi$ as well as on the type of altermagnetism and $\phi_{\rm L,R}$.  For $L_{\rm M}\gg\xi$,  only the spin-polarized ABSs  of the left JJ disperse with $\phi_{\rm L}$, while the spin-polarized ABSs of the right JJ remain dispersionless and intersect the ABSs of the left JJ at $\phi_{\rm R} = \phi_{\rm L}$; owing to the $2\pi$ periodicity of the spectrum, a similar intersection occurs at   $\phi_{\rm R} = 2\pi - \phi_{\rm L}$ [see Figs.\,\ref{Fig1}(b,c) and yellow/green circles]. The fact that ABSs of the left JJ  disperse only with $\phi_{\rm L}$ and those ones of the right JJ remain dispersionless occurs because the left and right JJs are weakly coupled due to $L_{\rm M}\gg\xi$. Interestingly, while  $d_{xy}$-wave altermagnetism splits  ABSs shifting them to distinct minima above and below $\phi_{\rm L}=\pi$, $d_{x^{2}-y^{2}}$-wave altermagnetism splits ABSs to higher and lower energies but maintaining the same minimum at $\phi=\pi$. This dependence in $d_{x^{2}-y^{2}}$-wave AMs makes ABSs to reach  and cross  zero energy, as in Fig.\,\ref{Fig1}(c).  Moreover, the finite size of the setup naturally induces a quasicontinuum, where, depending on the strength and type of altermagnetism, the spin-polarized ABSs can leak and/or touch the gap edges; for instance, under $d_{xy}$-wave altermagnetism,   ABSs with distinct   spin polarization touch the gap edges at $\phi_{\rm L}=0,2\pi$, while ABSs of the same polarization reach the gap edges at $\phi_{\rm L}=0,2\pi$ with   $d_{x^{2}-y^{2}}$-wave altermagnetism. Despite these differences between altermagnetic symmetries  at  $L_{\rm M}\gg\xi$, the energy spectrum is symmetric about $\phi_{\rm L}=\pi$ for any $\phi_{\rm R}$.

	When the middle superconductor is very short ($L_{\rm M}\lesssim\xi$), the spin-polarized ABSs of the left and right JJs undergo a strong hybridization that gives rise to spin-polarized Andreev molecules [see Figs.\,\ref{Fig1}(g,h)]. These spin-polarized Andreev molecules now constitute the subgap states of the strongly coupled JJs, and are, therefore, extremely sensible to variations of $\phi_{\rm L,R}$. In fact, $d_{xy}$-wave altermagnetism splits the hybridized ABSs into spin-polarized Andreev molecules along $\phi_{\rm L}$, developing minima below and above $\phi_{\rm L}^{*}$ [Fig.\,\ref{Fig1}(g)], with $\phi_{]rm L}^{*}$ being the value of $\phi_{\rm L}$ at which the Andreev molecule has a minimum without altermagnetism. This behavior enables JJs with $d_{xy}$-wave AMs to host Andreev molecules with  distinct polarizations at the same energies with a behavior that is asymmetric about $\phi_{\rm L}=\pi$. For $d_{x^{2}-y^{2}}$-wave altermagnetism, the  spin-polarized Andreev molecules are asymmetric about $\phi_{\rm L}=\pi$ and shifted in energy at $\phi_{\rm L}^{*}$; depending on the altermagnetic strength, they can reach and cross zero energy [see Fig.\,\ref{Fig1}(h)]. This leaves spin-polarized Andreev molecules with distinct spin polarizations crossing zero energy and at higher energies. The  zero-energy crossings of the Andreev molecules are robust unless fields that couple distinct spins are applied.  It is worth noting that the hybridized ABSs also leak into the quasicontinuum, but with subtle differences in each altermagnetic symmetry: The quasicontinuum of the JJ  with $d_{xy}$-wave altermagnetism is composed of mixed levels having the same and distinct polarizations, which implies that the spin-polarized Andreev molecule leaking into such quasicontinuum also hybridizes with the quasicontinuum levels of the same polarization [Fig.\,\ref{Fig1}(g)]. In contrast, the quasicontinuum of the JJ  with $d_{x^{2}-y^{2}}$-wave  altermagnetism contains spin-polarized subsets at distinct energies, which ensures that the spin-polarized Andreev molecules leaking into the quasicontinuum do not hybridize with the quasicontinuum levels [Fig.\,\ref{Fig1}(h)]. 
	
	The spin-polarized nature is also supported by the behavior of the spin polarization $\langle S_z \rangle$, which we obtain in the left JJ and show in Figs.\,\ref{Fig1}(d,e,i,j) as a function of $\phi_{\rm L,R}$ at $k_{y}\neq0$. Figures \ref{Fig1}(d) and \ref{Fig1}(e) show that $\langle S_z \rangle$ for $L_{\rm M}\gg\xi$ behaves as an odd function of $\phi_L$ for the $d_{xy}$-wave AMs ($\langle S_z(\phi_{\rm L})\rangle=-\langle S_z(-\phi_{\rm L})\rangle$), whereas it behaves as an even function for the $d_{x^2-y^2}$-wave AMs ($\langle S_z(\phi_{\rm L})\rangle=\langle S_z(-\phi_{\rm L})\rangle$). By close inspection, the profile of $\langle S_z \rangle$ aligns well with the spin-polarized nature of the ABSs shown in Figs.\,\ref{Fig1}(b,c) [see dash-dotted line in  Figs.\,\ref{Fig1}(d,e)]. In contrast, for $L_{\rm M}\lesssim\xi$, $\langle S_z \rangle$ acquires a distinct behavior that reflects opposite spin polarization along $\phi_{\rm L}=\phi_{\rm R}$: This implies that $\langle S_z(\phi_{\rm L},\phi_{\rm R})\rangle=\pm\langle S_z(-\phi_{\rm L},-\phi_{\rm R}) \rangle$ for the JJs with $d_{xy}$- and $d_{x^{2}-y^{2}}$-wave altermagnetism, reflecting the key role of $\phi_{\rm R}$. Still, $\langle S_z \rangle$ determines the spin polarization of the emergent Andreev molecules. Thus, Figs.\,\ref{Fig1}(d,e,i,j) not only support the spin polarized nature of Andreev molecules but also demonstrate the nonlocal phase control of spin polarization in coupled JJs.


	\section{Anomalous nonlocal Josephson currents}
	\label{sec:Anomalous_nonlocal_Josephson_currents}
	
\begin{figure}[!t]
		\centering
		\includegraphics[width=\linewidth]{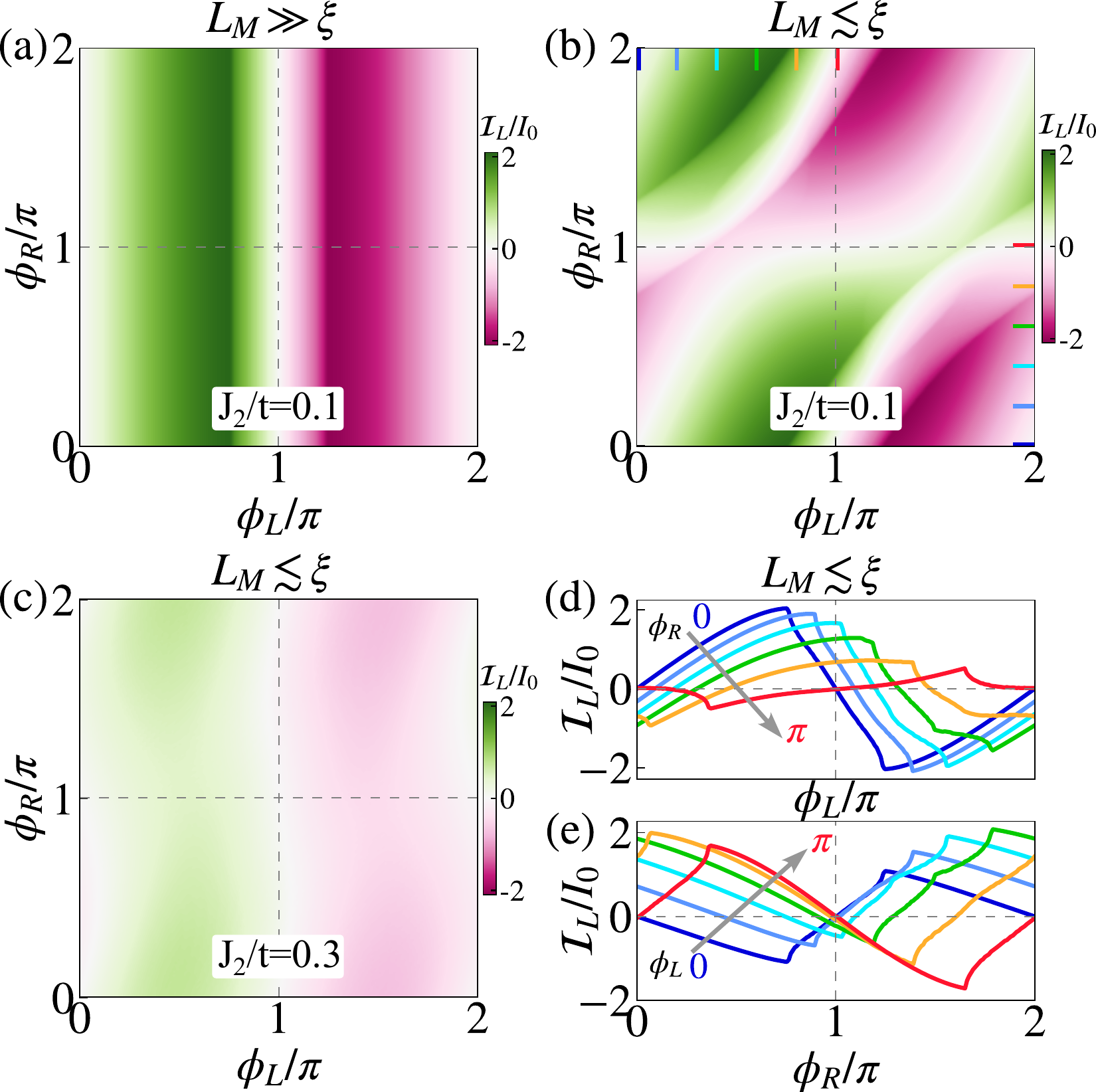}
		\caption{Josephson current ${\cal I}_{\rm L}$ across the left JJ as a function of $\phi_{\rm L,R}$ in the presence of $d_{x^{2}-y^{2}}$-wave altermagnetism. (a) ${\cal I}_{\rm L}$ for $L_{\rm M}=100a\gg\xi$ and $J_{2}/t=0.1$, and (b,c) for $L_{\rm M}=2a\lesssim\xi$ and $J_{2}/t=\{0.1,0.3\}$. (d,e) Line cuts of ${\cal I}_{\rm L}$ in (b) as a function of $\phi_{\rm L(R)}$ at distinct $\phi_{\rm R(L)}$ in steps of  $0.2\pi$  marked by color bars in (b). Rest of parameters are same in Fig.\,\ref{Fig1}}
		\label{Fig2} 
	\end{figure}

	Having shown the emergence of spin-polarized Andreev molecules, we now focus on the  phase-biased Josephson current flowing across the left JJ, obtained for each $k_{y}$  as  $I_{\rm L}(k_y, \phi_{\rm L}, \phi_{\rm R}) = (e/\hbar) \sum_{E_{n} <0} \partial E_{n}(k_y, \phi_{\rm L}, \phi_{\rm R})/\partial\phi_{\rm L}$ \cite{Beenakker:92,zagoskin,tanakaReview2024},   this provides a way to pinpoint the contributions of distinct energy levels \cite{furusaki1991dc,Furusaki_1999,kashiwaya2000,PhysRevB.96.205425,cayao2018andreev,PhysRevLett.123.117001,PhysRevB.104.L020501}. The total Josephson current flowing across the left JJ is then found as $\mathcal{I}_{\rm L}(\phi_{\rm L}, \phi_{\rm R}) = \int I_{\rm L}(k_y, \phi_{\rm L}, \phi_{\rm R}) dk_y$, which accounts for the  weight of all  transverse channels \cite{FukayaJPCM2025}. With this, in Fig.\,\ref{Fig2} we show $\mathcal{I}_{\rm L}(\phi_{\rm L}, \phi_{\rm R})$ as a function of $\phi_{\rm L,R}$ in the presence of $d_{x^{2}-y^{2}}$-wave altermagnetism for long ($L_{\rm M}\gg\xi$)  and short ($L_{\rm M}\lesssim\xi$) middle superconductors; for $d_{xy}$-wave altermagnetism, see Fig.\,\ref{FigE4} in App.\,\ref{app:JE_with_dxy_wave_AM}. When $L_{\rm M}\gg\xi$, the current $\mathcal{I}_{\rm L}(\phi_{\rm L}, \phi_{\rm R})$ is an odd function  and $2\pi$-periodic with respect to $\phi_{\rm L}$ but   insensitive to $\phi_{\rm R}$, namely, $\mathcal{I}_{\rm L}(\phi_{\rm L}, \phi_{\rm R})=-\mathcal{I}_{\rm L}(-\phi_{\rm L}, \phi_{\rm R})$ and $\mathcal{I}_{\rm L}(2\pi+\phi_{\rm L}, \phi_{\rm R})=\mathcal{I}_{\rm L}(\phi_{\rm L}, \phi_{\rm R})$  [Fig.\,\ref{Fig2}(a)]; these dependences   also appear  in regular JJs \cite{RevModPhys.76.411}.  This is an expected behavior consistent with the ABSs in Fig.\,\ref{Fig1}(c) since the left and right JJs are effectively uncoupled and the spin-polarized ABSs  of the right JJ  along with $\phi_{\rm R}$ do not affect the properties of the left JJ. 
	
	When $L_{\rm M}\lesssim\xi$, the hybridization of spin-polarized ABSs that leads to  spin-polarized Andreev molecules strongly affect the phase dependence of $\mathcal{I}_{\rm L}(\phi_{\rm L}, \phi_{\rm R})$ [see Fig.\,\ref{Fig2}(b-d)]. In particular, $\mathcal{I}_{\rm L}(\phi_{\rm L}, \phi_{\rm R})$ is now strongly dependent on $\phi_{\rm R}$, developing negative/positive values for $\phi_{\rm L}<\pi$ and $\phi_{\rm L}>\pi$, which depend on the strength of altermagnetism and are achieved after passing through zero values in both types of the studied AMs [see Figs.\,\ref{Fig2}(b,c) and Figs.\,\ref{FigE4}(b,c)].  
	Interestingly, $\mathcal{I}_{\rm L}(\phi_L, \phi_R)\neq0$ at $\phi_{\rm L}=n\pi$, with $n\in\mathbb{Z}$, and its oddness is now reflected by the simultaneous action of both phases $\phi_{\rm L,R}$ as  $\mathcal{I}_{\rm L}(\phi_L, \phi_R)=-\mathcal{I}_{\rm L}(-\phi_L, -\phi_R)$; this occurs in $d_{x^{2}-y^{2}}$- and $d_{xy}$-wave altermagnetism [see Figs.\,\ref{Fig2}(b,c) and Figs.\,\ref{FigE4}(b,c)]. As consequence of these 
	properties is that the current flowing across the left junction develops  $\phi_{0}$- and $\pi$-junction behaviors, as a function of either $\phi_{\rm L}$ or $\phi_{\rm R}$ [see Figs.\,\ref{Fig2}(d,e)]. It is thus possible to have $\mathcal{I}_{\rm L}(\phi_{\rm L}=0, \phi_{\rm R}\neq0)\neq0$ or $\mathcal{I}_{\rm L}(\phi_{\rm L}+\pi, \phi_{\rm R})=-\mathcal{I}_{\rm L}(\phi_{\rm L}, \phi_{\rm R})$ for a  $\phi_{L0}$ junction or $\pi$ junction, respectively. JJs with  $d_{xy}$-wave altermagnetism can also host similar  junction profiles [see Figs.\,\ref{FigE4}(d,e)].
	
	The strong dependence of the current flowing across the left JJ by varying the phase difference across the right JJ signals the emergence of the nonlocal Josephson effect. This effect in the left JJ is entirely determined by the spin-polarized Andreev molecules shown in Figs.\,\ref{Fig1}(g,h), with current signals that can be directly tied to the respective energy branch. For instance, the large values of the current for $\phi_{\rm L}\in(0,\pi)$ and $\phi_{\rm L}\in(\pi,2\pi)$ in JJs with $d_{x^{2}-y^{2}}$-wave altermagnetism are predominantly  carried by Andreev molecules of one type of spin polarization within the gap, while outside   it the opposite polarization also contributes [Fig.\,\ref{Fig1}(h)]. In contrast, the arc surface region with smaller current intensities for $\phi_{\rm L}\in(\pi,2\pi)$ corresponds to a regime where the Andreev molecules with distinct polarizations cross zero energy [Fig.\,\ref{Fig1}(h)], leaving Andreev molecules with the same polarization but opposite contribution that end up cancelling out; thus, the values of the current in this case are dominated by the quasicontinuum. In a similar way, one can identify the contribution of spin-polarized Andreev molecules on the nonlocal Josephson effect in coupled JJs with $d_{xy}$-wave altermagnetism [see Figs.\,\ref{FigE4}(b,c) in App.\,\ref{app:JE_with_dxy_wave_AM}]. We can therefore conclude that the formation of spin-polarized Andreev molecules enables the emergence of a nonlocal Josephson effect determining supercurrents with anomalous profiles.

	\begin{figure}[!t]
		\centering
		\includegraphics[width=1.0\linewidth]{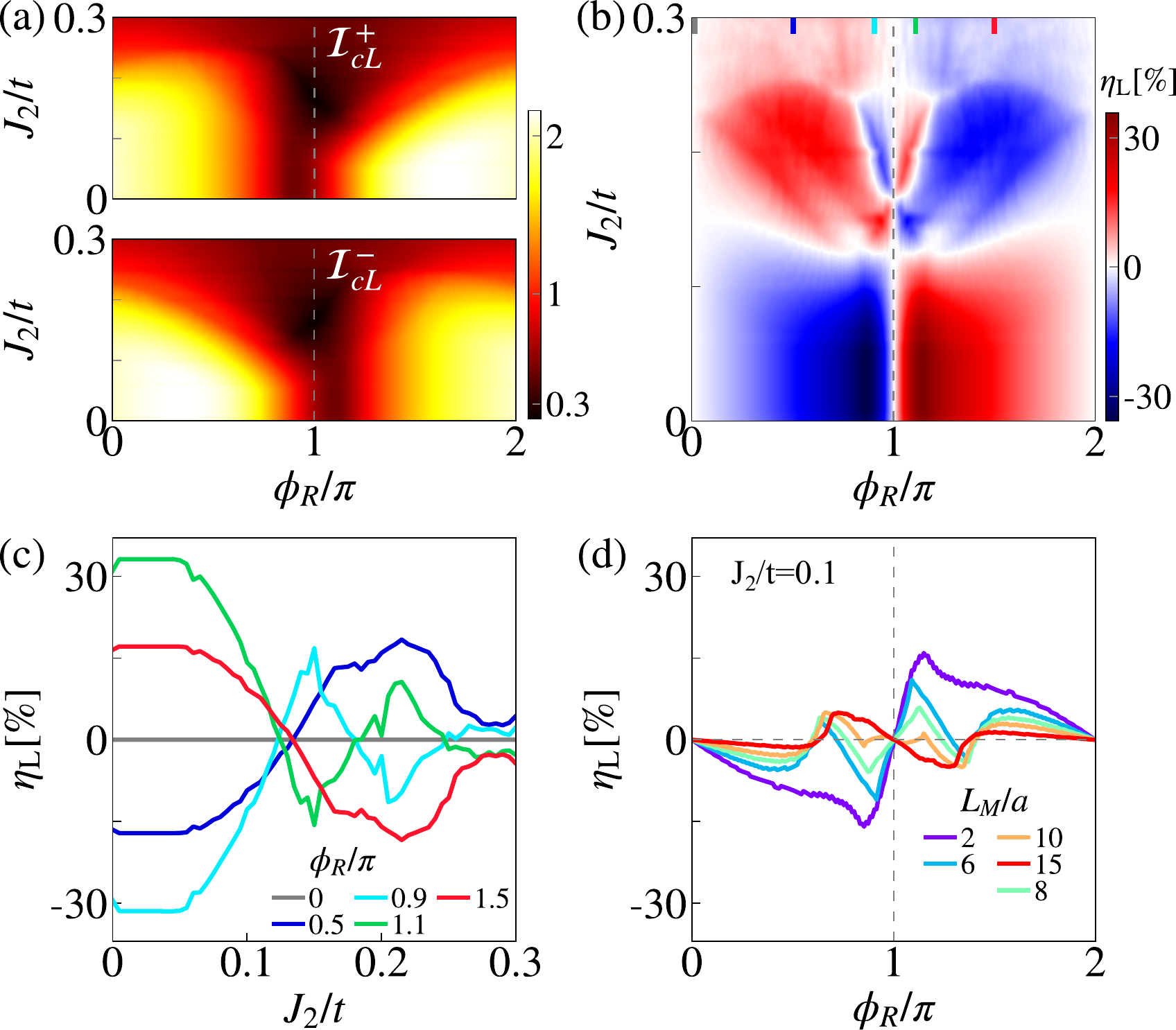}
		\caption{(a) Critical currents  $\mathcal{I}_{\rm cL}^{\pm}$ as a function of  $\phi_{\rm R}$ and $J_2$ under $d_{x^{2}-y^{2}}$-wave altermagnetism. (b) Quality factor  $\eta_{\rm L}$  as a function of  $J_2$ and $\phi_{\rm R}$. (c,d) Line cuts of (b) for distinct values of $J_{2}$ and  $\phi_{\rm R}$. The values of $\phi_{\rm R}$ for the line cuts in (c) are marked by color bars in (b). Parameters: $L_{\rm M} = 2a$ and the rest is the same as in Fig.\,\ref{Fig1}.}
		\label{Fig3} 
	\end{figure}

	\section{Nonlocal Josephson diode effect}
	\label{sec:Nonlocal_JDE}
	
	The   behavior of the nonlocal Josephson current $\mathcal{I}_{\rm L}(\phi_{\rm L}, \phi_{\rm R})$   also reveals  its nonreciprocity since the maximum and minimum currents are not equal, $\mathcal{I}_{\rm cL}^{+}(\phi_{\rm R})\neq\mathcal{I}_{\rm cL}^{-}(\phi_{\rm R})$, where
	$\mathcal{I}_{\rm cL}^{\pm}(\phi_{\rm R})={\rm max}_{\phi_{\rm L}}[\pm\mathcal{I}_{\rm L}(\phi_L, \phi_{\rm R})]$.  Using these definitions, in Fig.\,\ref{Fig3}(a) we show $\mathcal{I}_{\rm cL}^{\pm}(\phi_{\rm R})$ as a function of $J_{2}$ and $\phi_{\rm R}$ for a very short middle region $L_{\rm M}=2a$ in the presence of $d_{x^{2}-y^{2}}$-wave altermagnetism. 	Figure \ref{Fig3}(a) shows that $\mathcal{I}_{\rm cL}^{+}(\phi_{\rm R})\neq\mathcal{I}_{\rm cL}^{-}(\phi_{\rm R})$ for any $\phi_{\rm R}$, except $\phi_{\rm R}=n\pi$, with $n\in\mathbb{Z}$.  The nonreciprocity in the nonlocal critical currents signals  the emergence of the nonlocal Josephson diode effect. While $\mathcal{I}_{\rm cL}^{\pm}(\phi_{\rm R})$   remains sizable for $J_{2}$ below $\Delta$, strong altermagnetic strengths tend to reduce the critical currents; in JJs with $d_{x^{2}-y^{2}}$-wave altermagnetism [Fig.\,\ref{Fig3}(a)], the currents support stronger values of altermagnetic fields than in  JJs with $d_{xy}$-wave AMs [see App.\,\ref{app:diode_effect_with_dxy_wave_AM} and Fig.\,\ref{FigE5}(a)]. 
	
	To assess the efficiency of the nonlocal Josephson diode effect,  we obtain its quality factor as $\eta_{\rm L}(\phi_{\rm R})=[\mathcal{I}_{\rm cL}^{+}(\phi_{\rm R}) - \mathcal{I}_{\rm cL}(\phi_{\rm R})^{-}]/[\mathcal{I}_{\rm cL}^{+}(\phi_{\rm R}) + \mathcal{I}_{\rm cL}^{-}(\phi_{\rm R})]$ and show it in Fig.\,\ref{Fig3}(b) as a function of $J_{2}$ and $\phi_{\rm R}$ for JJs with $d_{x^{2}-y^{2}}$-wave altermagnetism, while in  Fig.\,\ref{FigE5}(b) for the $d_{xy}$-wave AM case. The first observation  is that $\eta_{\rm L}$ strongly depends on $\phi_{\rm R}$ as well as on the strength and type of altermagnetism. For $\phi_{\rm R}= n\pi$, with $n\in\mathbb{Z}$, the quality factor vanishes since $\mathcal{I}_{\rm cL}^{+}=\mathcal{I}_{\rm cL}^{-}$; in this case, the Andreev spectrum is symmetric around $\phi_{\rm L}=\pi$. Away from these phase values, $\eta_{\rm L}$  reach efficiencies on the order of $30\%$ and notably  develops an odd function dependence with  $\phi_{\rm R}$ as a clear manifestation of  the nonlocal Josephson   effect; this also means that the diode's polarity ${\rm sgn}(\eta_{\rm L})$  can be controlled   nonlocally by $\phi_{\rm R}$ [see Fig.\,\ref{Fig3}(c) and Fig.\,\ref{FigE5}(c)]. Furthermore, the diode's polarity can be controlled by the strength   of altermagnetism, as demonstrated in Figs.\,\ref{Fig3}(c,d) and Figs.\,\ref{FigE5}(c,d) at fixed $\phi_{\rm R}$. Under $d_{x^{2}-y^{2}}$-wave altermagnetism, changing the   diode's polarity requires slightly stronger altermagnetic fields than under $d_{xy}$-wave altermagnetism but the former supports larger efficiencies at large fields. Nevertheless, in both types of AMs, reaching measurable  nonlocal Josephson diode efficiencies requires the presence of spin-polarized Andreev molecules, which is only attained    for short middle superconductors, as demonstrated in Figs.\,\ref{Fig3}(d) and \ref{FigE5}(d).

	\section{Conclusions}
	\label{sec:conclusions}
	
	In conclusion, we have demonstrated the formation of spin-polarized Andreev molecules in weakly coupled Josephson junctions   with zero net magnetization by utilizing altermagnetism. We showed that these spin-polarized Andreev molecules give rise to  phase-biased nonlocal Josephson currents  capable of developing $\phi_{0}$- and $\pi$-junction behaviors, which are tunable by the type and strength of altermagnetism. Moreover, we found that these nonlocal Josephson currents exhibit nonreciprocal  critical currents, originating a nonlocal Josephson diode effect whose polarity is controllable nonlocally by the superconducting phase difference across the right Josephson junction as well as by the strength of altermagnetism. These predictions are applicable to higher angular momentum altermagnets, such as $g$ and $i$ wave, and   offer a very likely nonlocal phase control of   recently reported   altermagnet-induced phenomena, such as   exotic Cooper pairs \cite{Maeda2025, chakraborty2024,khodas2025strain,PhysRevB.111.054520,parshukov2025,mazin2025notes,fu2025light,fu2025floquet,Yokoyama25floquet,mukasa2025finite,heinsdorf2025, monkman2026persistent,ChangPRB2025,khodas2026pUM} and flat bands \cite{lu2025subgap,liu2025FFLOBFS,fukaya2026crossed}. 
	We emphasize that the diverse phenomena reported here are not independent effects, but rather distinct macroscopic manifestations of a single microscopic origin. By combining a vanishing net magnetization with momentum-dependent spin splitting, altermagnetism fundamentally spin polarizes the hybridized Andreev molecules. This resulting molecular spectrum serves as the unified engine driving the anomalous nonlocal currents, the junction-character transitions, and the tunable diode polarity.  Our results position altermagnetism as a foundational building block for engineering spin-polarized Andreev molecules and achieving anomalous Josephson states with nonlocal spintronic functionalities.

	\begin{acknowledgements}
	We acknowledge financial support from the G\"{o}ran Gustafsson Foundation (Grant No. 2216), from the Swedish Research Council (Vetenskapsr{\aa}det Grant No. 2021-04121), and from the Olle Engkvist Foundation (Grant No.  243-1026). The computations were enabled by resources provided by the National Academic Infrastructure for Supercomputing in Sweden (NAISS), partially funded by the Swedish Research Council through Grant Agreement No. 2022-06725.
	\end{acknowledgements}


	\appendix

	\section{Molecular behavior of the hybridized Andreev states}
	\label{app:Molecular behavior of the hybridized Andreev states}

\begin{figure}[!t]
	\centering
	\includegraphics[width=1.0\linewidth]{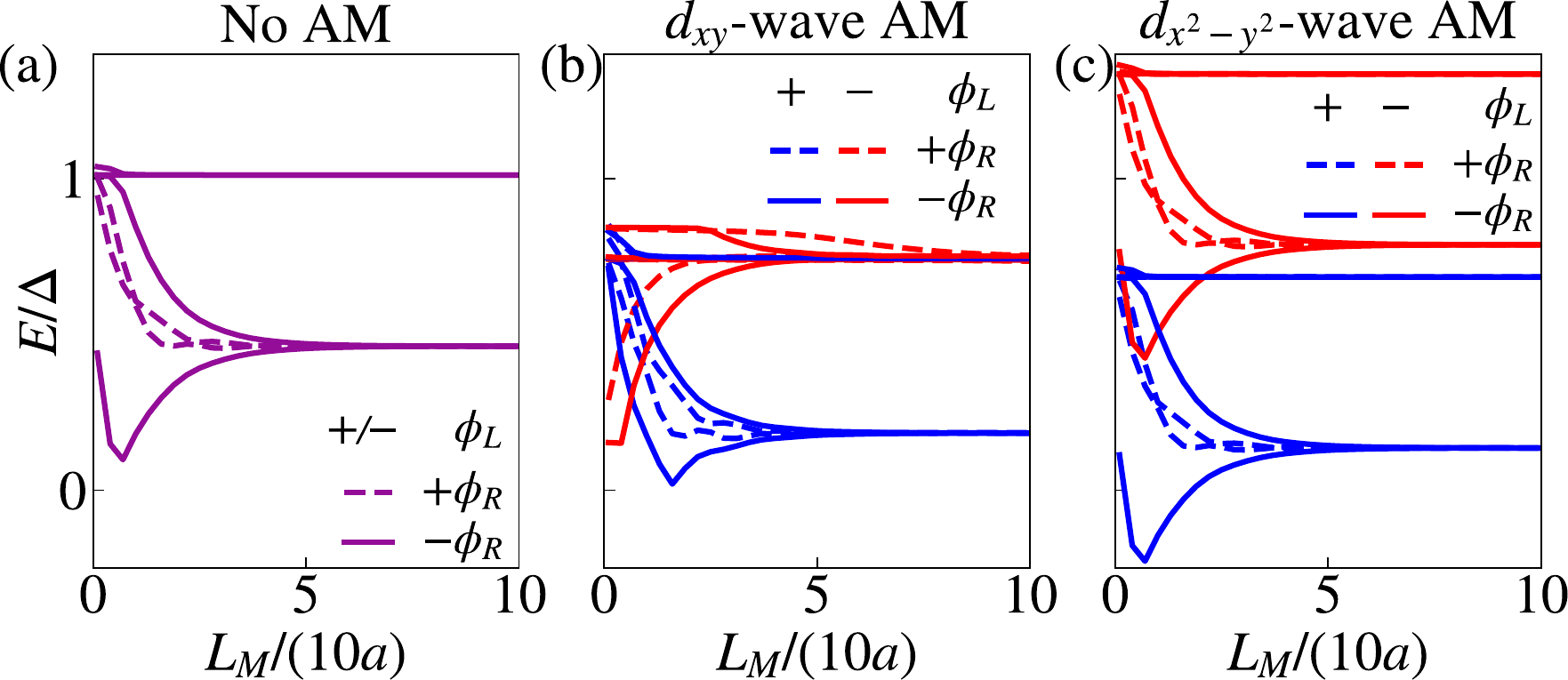}
	\caption{Lowest four positive   levels of each  $\pm$  sector  as a function of  $L_{\rm M}$ in the absence [panel (a)] and presence of altermagnetism  [Panels (b), (c)] at $\phi_{\rm R} = 0.7\pi$ and $k_y = 0.1\pi$.   The dashed and solid curves correspond to $\phi_{\rm L} = 0.7\pi = \phi_{\rm  R}$ and $\phi_{\rm L} = 1.3\pi =  -\phi_{\rm  R}$ respectively.    Parameters: $J_{1,2} = 0$ in panel (a), $J_1 = 0.1t$ and $J_2 = 0$ in panel (b), $J_1 = 0$ and $J_2 = 0.1t$ in panel (c), and the rest   as in Fig.\,\ref{Fig1}.}
	\label{FigE2} 
\end{figure}

	The molecular nature of the spin-polarized ABSs can be directly seen in the dependence of the energy levels on $L_{\rm M}$,   presented in  Fig.\,\ref{FigE2} at $\phi_{\rm L}=0.7\pi=\phi_{\rm R}$ and $\phi_{\rm L}=1.3\pi\neq\phi_{\rm R}$ for   altermagnetic strengths corresponding to Fig.\,\ref{Fig1}. Without AMs, the energy levels are spin degenerate  but already here the levels   develop an interesting behavior with $L_{\rm M}$ [see Fig.\,\ref{FigE2}(a)]: At long $L_{\rm M}$, each JJ hosts  ABSs  that are degenerate but begin to split at $L_{\rm M}\sim 50a$, followed by a behavior that depends on $\phi_{\rm L,R}$. For $\phi_{\rm L}=1.3\pi\neq\phi_{\rm R}$, the energy levels below $L_{\rm M}\sim 50a$ form bonding and antibonding molecular states, with a  clear minimum and an overall energy dependence that resembles the interatomic potential describing molecular formation [Fig.\,\ref{FigE2}(a)]. For $\phi_{\rm L}=0.7\pi=\phi_{\rm R}$, the splitting between ABSs induced by the hybridization   is weaker below $L_{\rm M}\sim 50a$, ultimately pushing both levels into the quasicontinuum.  Interestingly, in the presence of altermagnetism, ABSs are split in spin and, interestingly, such spin-polarized ABSs still develop the molecular formation profile for $\phi_{\rm L}=1.3\pi\neq\phi_{\rm R}$ and in both types of altermagnetism [see Figs.\,\ref{FigE2}(b,c)]. This demonstrates the emergence of spin-polarized Andreev molecules.

\section{Localization of Andreev molecules}
\label{app:localization of Andreev molecules}

\begin{figure}[!h]
	\centering
	\includegraphics[width=1.0\linewidth]{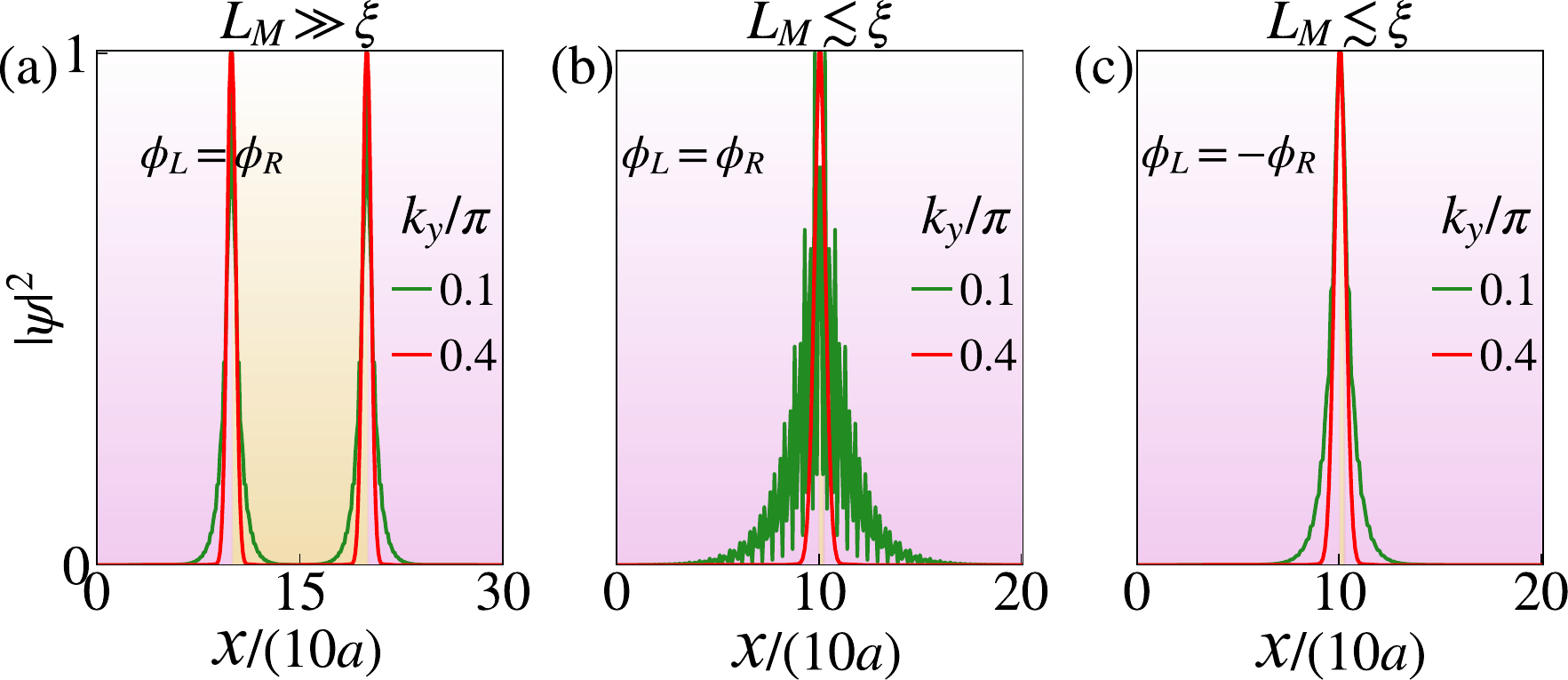}
	\caption{Wave function probability density of the first positive state 
		for a JJ with $d_{x^{2}-y^{2}}$-wave altermagnetism as a function of space  for $L_M = 100a$ [panel (a)] and $L_M=2a$ [panels (b),(c)], in both cases with $\phi_{\rm R}=0.7\pi$.   The green and the red curves indicate two different values of $k_y$. Parameters: $J_1 = 0$ and $J_2 = 0.1t$; the rest is the same as in Fig.\,\ref{Fig1}.}
	\label{FigE1} 
\end{figure}

The hybridization of spin-polarized ABSs in the JJ setup of Fig.\,\ref{Fig1}(a) is due to the reduced length of the middle superconductor, which leads to a strong coupling between left and right JJs. To see the real-space behavior of such hybridization, Fig.\,\ref{FigE1} shows the space dependence of the wave function probability density of  the  first positive state for   $L_{\rm M} = 100a$ and  $L_{\rm M} = 2a$ for distinct $\phi_{\rm L,R}$.  When $L_{\rm M} = 100a$, the coupling between left and right JJs is very  weak, such that the ABSs localized at each JJ do not hybridize [Fig.\,\ref{FigE1}(a)]: As seen, the ABSs are strongly localized at each JJ, with their wave functions decaying toward the bulk of the middle superconductor that depends on $k_{y}$ but without developing any spatial overlap. In contrast, for  $L_{\rm M} = 2a$, the wave functions of the ABSs of the left and right JJs strongly hybridize, with a profile that resembles that of an  ABS  localized at a single interface. The wave function profile of this spin-polarized Andreev molecule depends on $k_{y}$ and $\phi_{\rm L,R}$, which can influence the decay itself  and its oscillations.

\section{Nonlocal Josephson currents in junctions with $d_{xy}$-wave altermagnetism}
\label{app:JE_with_dxy_wave_AM}

\begin{figure}[!t]
		\centering
		\includegraphics[width=1.0\linewidth]{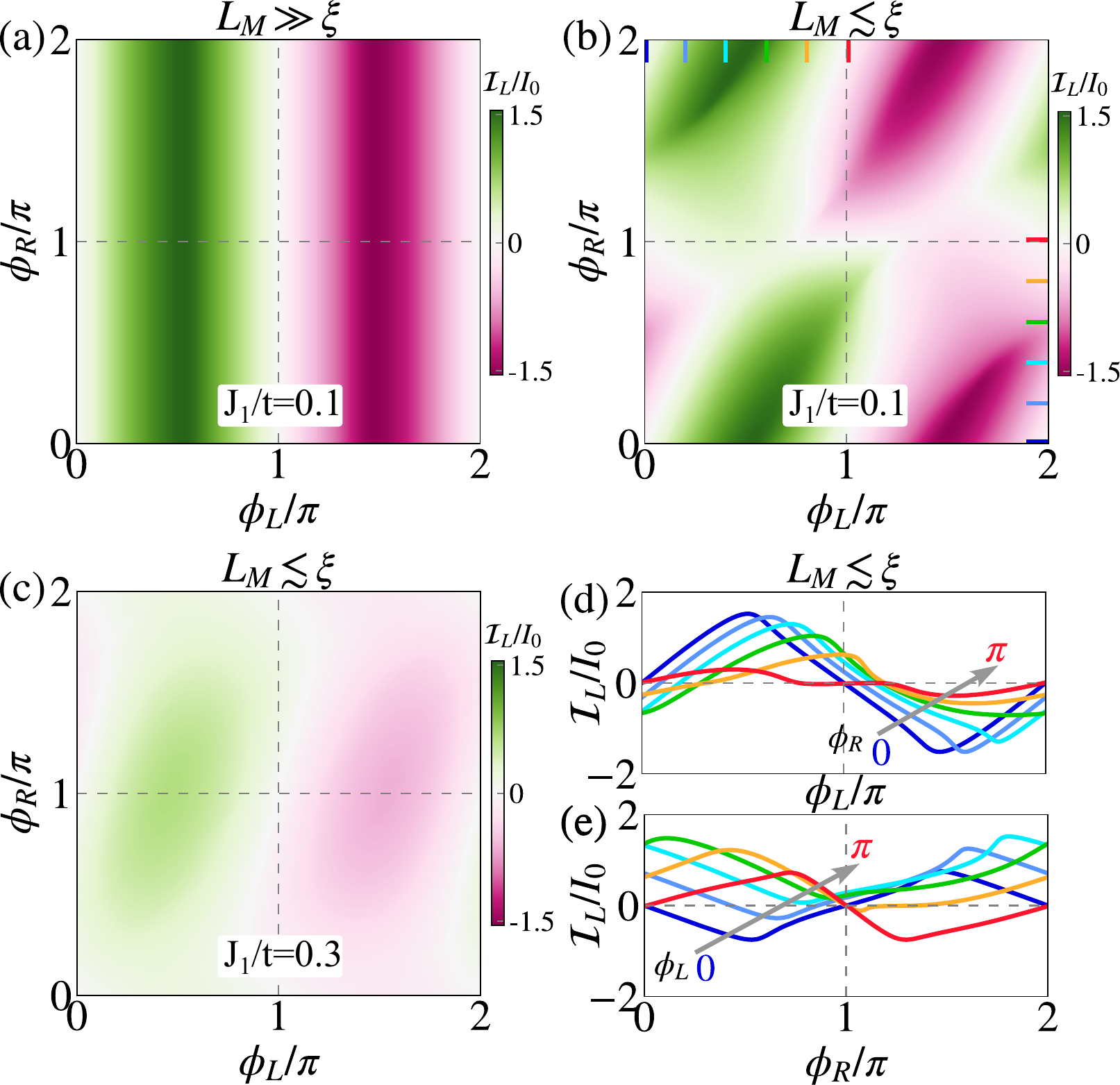}
		\caption{The Josephson current ${\cal I}_{\rm L}$ across the left JJ as a function of $\phi_{\rm L,R}$ in the presence of $d_{xy}$-wave altermagnetism.  ${\cal I}_{\rm L}$ for (a) $L_{\rm M}=100a\gg\xi$ and $J_{1}/t=0.1$, and (b), (c) for $L_{\rm M}=2a\lesssim\xi$ and $J_{1}/t=\{0.1,0.3\}$. (d), (e) Line cuts of ${\cal I}_{\rm L}$ in panel (b) as a function of $\phi_{\rm L(R)}$ at distinct $\phi_{\rm R(L)}$ in steps of  $0.2\pi$ marked by color bars in panel (b). Parameters are same in Fig.\,\ref{Fig1}. }
		\label{FigE4} 
	\end{figure}

 In the case of $d_{xy}$-wave altermagnetism, the total Josephson current flowing across the left JJ, $\mathcal{I}_{\rm L}(\phi_{\rm L}, \phi_{\rm R})$, is presented in Figs.\,\ref{FigE4}(a,b) as a function of $\phi_{\rm L,R}$  for long ($L_{\rm M}\gg\xi$)  and short ($L_{\rm M}\lesssim\xi$) middle superconductors at $J_{1}/t=0.1$. Figure \ref{FigE4}(c) shows $\mathcal{I}_{\rm L}(\phi_{\rm L}, \phi_{\rm R})$ versus $\phi_{\rm L,R}$ for $J_{1}/t=0.3$, while Figs.\,\ref{FigE4}(d,e) shows line cuts of  Fig.\,\ref{FigE4}(b) as a function of $\phi_{\rm L,R}$ at distinct $\phi_{\rm R,L}$. For $L_{\rm M}\gg\xi$, Fig.\,\ref{FigE4}(a) reveals that  $\mathcal{I}_{\rm L}(\phi_{\rm L}, \phi_{\rm R})$ is an odd function of $\phi_{\rm R}$ but independent of $\phi_{\rm R}$; this is because   the left and right JJs are effectively uncoupled (very weakly coupled) and the spin-polarized ABSs in each JJ do not hybridize.    When $L_{\rm M}\lesssim\xi$,  $\mathcal{I}_{\rm L}(\phi_{\rm L}, \phi_{\rm R})$ exhibits large intensities that strongly depend on both $\phi_{\rm L,R}$ and $J_{1}$ [see Fig.\,\ref{FigE4}(b-d)]. First of all, $\mathcal{I}_{\rm L}(\phi_{\rm L}, \phi_{\rm R})$ is an odd function under both $\phi_{\rm L,R}$ and develops nonzero values at $\phi_{\rm L}=n\pi$ for $\phi_{\rm R}\neq n\pi$ [Fig.\,\ref{FigE4}(b)], similar to JJs with $d_{x^{2}-y^{2}}$-wave altermagnetism shown in Figs.\,\ref{Fig2}(b,d,e).  As a result, the current across the left JJ can develop  $\phi_{\rm L0}$-junction behavior as a function of $\phi_{\rm L,R}$ but   $\pi$ behavior only as a function of $\phi_{\rm R}$ [Figs.\,\ref{Fig2}(b,d,e)], hence establishing similarities and differences with respect to JJs with  $d_{x^{2}-y^{2}}$-wave altermagnetism. The different behaviors can be controlled by the strength of $d_{xy}$-wave altermagnetism [Fig.\,\ref{Fig2}(c)].
	
	The intriguing behavior of the anomalous Josephson current $\mathcal{I}_{\rm L}(\phi_{\rm L}, \phi_{\rm R})$ discussed above can be directly traced back to the behavior of the spin-polarized Andreev molecules shown in Fig.\,\ref{Fig1}(g). For instance, for $\phi_{\rm R}=0.7\pi$ in Fig.\,\ref{Fig1}(g), the Andreev molecules inside the gap with positive spin polarization produce a negative supercurrent roughly below $\phi_{\rm L}<0.7$, while above such a phase, the Andreev molecule with opposite  polarization dominates and gives a positive supercurrent [Fig.\,\ref{FigE4}(b)]; this  explains    the transition from a red to green color in the intensities of the current in Fig.\,\ref{FigE4}(b) [see also region between orange and green  horizontal markers in Fig.\,\ref{FigE4}(b)]. The absence of a $\pi$-junction behavior as a function of $\phi_{\rm L}$ can also be explained with the behavior of the Andreev molecules: Andreev molecules with both polarizations always share the contribution within the gap and  $\phi_{\rm L}\in(0,2\pi)$ since $d_{xy}$-wave altermagnetism shifts the subgap levels only in $\phi_{\rm L}$. This makes it challenging to  accommodate a single Andreev molecule giving a negative current for JJs with $d_{xy}$-wave altermagnetism. This is different from what occurs in  $d_{x^{2}-y^{2}}$-wave altermagnetism [Figs.\,\ref{Fig2}(b,d)] since   therein the altermagnetic field shifts the energy levels in energy, which can leave a single Andreev molecule  with a phase dependence supporting a $\pi$ junction. In sum, the nonlocal Josephson currents across the left JJ under $d_{xy}$-wave altermagnetism  can develop $\phi_{\rm L0}$- and $\pi$-junction behaviors controlled by the superconducting phases and altermagnetic strength.

\section{Nonlocal Josephson diode effect in junctions with $d_{xy}$-wave altermagnetism}
\label{app:diode_effect_with_dxy_wave_AM}

	\begin{figure}[!t]
	\centering
	\includegraphics[width=1.0\linewidth]{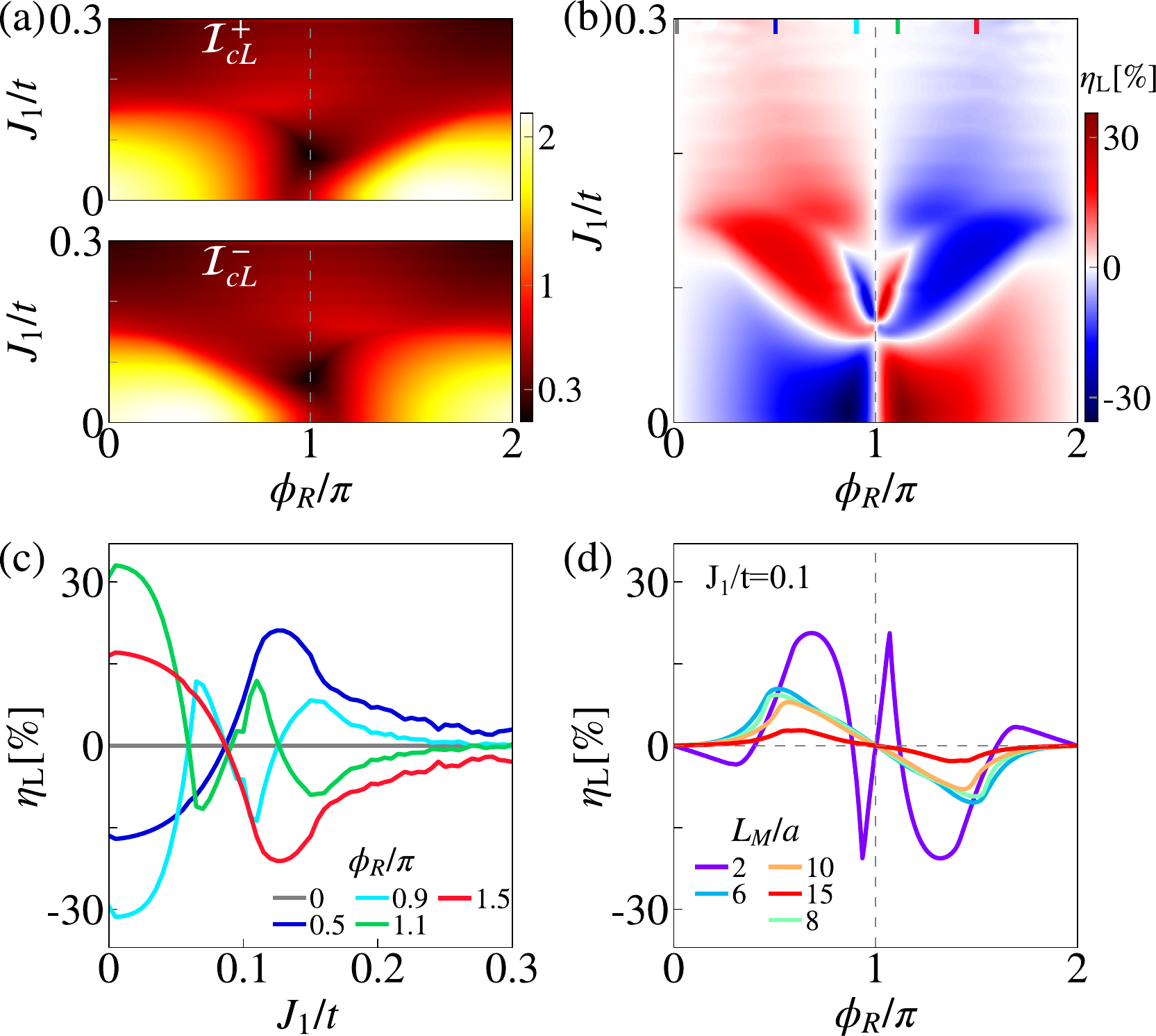}
	\caption{(a) Critical currents  $\mathcal{I}_{\rm cL}^{\pm}$ as a function of 
		$\phi_{\rm R}$ and $J_1$ under $d_{xy}$-wave altermagnetism. (b) Quality factor  $\eta_{\rm L}$ as a function of  $J_1$ and $\phi_R$. (c), (d) Line cuts of panel (b) for distinct values of $J_{1}$ and  $\phi_{\rm R}$. The values of $\phi_{\rm R}$ for the line cuts in panel (c) are marked by color bars in panel (b). Parameters: $L_M = 2a$ and the rest is the same as in Fig.\,\ref{Fig1}.}
	\label{FigE5} 
\end{figure}
	
	The anomalous nonlocal Josephson currents flowing across the left junction of Fig.\,\ref{FigE4} also exhibit regimes with distinct critical currents $\mathcal{I}_{\rm cL}^{+}(\phi_{\rm R})\neq\mathcal{I}_{\rm cL}^{-}(\phi_{\rm R})$, hence leading to a nonlocal Josephson diode effect under $d_{xy}$-wave altermagnetism characterized by a nonzero quality factor $\eta_{\rm L}(\phi_{\rm R})\neq0$. To understand this behavior, in Figs.\,\ref{FigE5}(a,b) we present $\mathcal{I}_{\rm cL}^{\pm}(\phi_{\rm R})$ and $\eta_{\rm L}(\phi_{\rm R})$ as functions of $\phi_{\rm  R}$ and $J_{1}$ for a short middle superconductor ($L_{\rm M}\lesssim\xi$); in  Figs.\,\ref{FigE5}(c,d) we further show $\eta_{\rm L}(\phi_{\rm R})$ as a function of $J_{1}$ for distinct $\phi_{\rm R}$ and as a function of $\phi_{\rm R}$ for different values of $L_{\rm M}$. First, the critical currents $\mathcal{I}_{\rm cL}^{\pm}(\phi_{\rm R})$ exhibit values that do not follow the same dependence as a function of $\phi_{\rm R}$ and $J_{1}$, and with a particular asymmetric profile about $\phi_{\rm R}=\pi$ [see Fig.\,\ref{FigE5}(a)]. By acquiring distinct values, the critical currents become nonreciprocal and signal the emergence of a nonlocal Josephson diode effect in coupled JJs with $d_{xy}$-wave altermagnetism. This diode effect is characterized by a finite quality factor $\eta_{\rm L}$ in Fig.\,\ref{FigE5}(b), which acquires an odd function dependence in $\phi_{\rm R}$ that serves as a knob to control the diode's polarity. Notably, the nonlocal diode's polarity can also be controlled by the strength of $d_{xy}$-wave altermagnetism, which can be seen by noting the change in the sign of $\eta_{\rm L}$ with $J_{1}$ in Fig.\,\ref{FigE5}(b) [see also Fig.\,\ref{FigE5}(c)]. This implies that $d_{xy}$-wave altermagnetism acts a switcher to turn on and off the nonlocal Josephson diode effect, in the same way as $d_{x^{2}-y{^2}}$-wave altermagnetism shown in Figs.\,\ref{Fig3}(b,c). While there exist some similarities in how the diode's quality factor responds to both types of altermagnetic symmetries, the  field strengths needed for manipulating the  polarity are lower in  coupled JJs with  $d_{xy}$-wave altermagnetism, but larger efficiencies are supported with  $d_{x^{2}-y{^2}}$-wave altermagnetism.

	\bibliography{biblio}

@Article{Cayao2020odd,
author={Cayao, Jorge
and Triola, Christopher
and Black-Schaffer, Annica M.},
title={Odd-frequency superconducting pairing in one-dimensional systems},
journal={Eur. Phys. J. Special Topics},
year={2020},
month={Feb},
day={01},
volume={229},
number={4},
pages={545-575},
issn={1951-6401},
doi={10.1140/epjst/e2019-900168-0},
url={https://doi.org/10.1140/epjst/e2019-900168-0}
}

@article{ma2021multifunctional,
	title={Multifunctional antiferromagnetic materials with giant piezomagnetism and noncollinear spin current},
	author={Ma, Hai-Yang and Hu, Mengli and Li, Nana and Liu, Jianpeng and Yao, Wang and Jia, Jin-Feng and Liu, Junwei},
	journal={Nat. Commun.},
	volume={12},
	number={1},
	pages={2846},
	year={2021},
	publisher={Nature Publishing Group UK London},
	url={https://doi.org/10.1038/s41467-021-23127-7}
}

@article{yang2026topological,
	title={Topological altermagnetic {J}osephson junctions},
	author={Yang, Grant ZX and Sun, Zi-Ting and Xie, Ying-Ming and Law, KT},
	journal={npj Quantum Materials},
	volume={11},
	pages={45},
	year={2026},
	publisher={Nature Publishing Group UK London},
	url={https://doi.org/10.1038/s41535-026-00874-8}
}

@article{song2025altermagnets,
	title={Altermagnets as a new class of functional materials},
	author={Song, Cheng and Bai, Hua and Zhou, Zhiyuan and Han, Lei and Reichlova, Helena and Dil, J Hugo and Liu, Junwei and Chen, Xianzhe and Pan, Feng},
	journal={Nat. Rev. Mater.},
	volume={10},
	number={6},
	pages={473--485},
	year={2025},
	publisher={Nature Publishing Group UK London},
	url ={https://doi.org/10.1038/s41578-025-00779-1}
}

@article{PhysRevB.109.245133,
  title = {Strong nonlocal tuning of the current-phase relation of a quantum dot based {A}ndreev molecule},
  author = {Kocsis, M{\'a}ty{\'a}s and Scher{\"u}bl, Zolt{\'a}n and F{\"u}l{\"o}p, Gerg{\H{o}} and Makk, P{\'e}ter and Csonka, Szabolcs},
  journal = {Phys. Rev. B},
  volume = {109},
  issue = {24},
  pages = {245133},
  numpages = {15},
  year = {2024},
  month = {Jun},
  publisher = {American Physical Society},
  doi = {10.1103/PhysRevB.109.245133},
  url = {https://link.aps.org/doi/10.1103/PhysRevB.109.245133}
}

@article{PhysRevB.109.L081405,
  title = {Enhancing the {J}osephson diode effect with {M}ajorana bound states},
  author = {Cayao, Jorge and Nagaosa, Naoto and Tanaka, Yukio},
  journal = {Phys. Rev. B},
  volume = {109},
  issue = {8},
  pages = {L081405},
  numpages = {7},
  year = {2024},
  month = {Feb},
  publisher = {American Physical Society},
  doi = {10.1103/PhysRevB.109.L081405},
  url = {https://link.aps.org/doi/10.1103/PhysRevB.109.L081405}
}

@article{zhang2022,
	title={General theory of {J}osephson diodes},
	author={Zhang, Yi and Gu, Yuhao and Li, Pengfei and Hu, Jiangping and Jiang, Kun},
	journal={Phys. Rev. X},
	volume={12},
	number={4},
	pages={041013},
	year={2022},
	publisher={APS},
	url={https://doi.org/10.1103/PhysRevX.12.041013}
}

@article{PhysRevResearch.6.L022002,
	title = {Tuning the {J}osephson diode response with an ac current},
	author = {Seoane Souto, Rub\'en and Leijnse, Martin and Schrade, Constantin and Valentini, Marco and Katsaros, Georgios and Danon, Jeroen},
	journal = {Phys. Rev. Res.},
	volume = {6},
	issue = {2},
	pages = {L022002},
	numpages = {7},
	year = {2024},
	month = {Apr},
	publisher = {American Physical Society},
	doi = {10.1103/PhysRevResearch.6.L022002},
	url = {https://link.aps.org/doi/10.1103/PhysRevResearch.6.L022002}
}

@article{davydova2022universal,
  title={Universal {J}osephson diode effect},
  author={Davydova, Margarita and Prembabu, Saranesh and Fu, Liang},
  journal={Sci. Adv.},
  volume={8},
  number={23},
  pages={eabo0309},
  year={2022},
  url={https://www.science.org/doi/full/10.1126/sciadv.abo0309}
}

@article{PhysRevLett.99.067004,
  title = {Proposed Design of a {J}osephson Diode},
  author = {Hu, Jiangping and Wu, Congjun and Dai, Xi},
  journal = {Phys. Rev. Lett.},
  volume = {99},
  issue = {6},
  pages = {067004},
  numpages = {4},
  year = {2007},
  month = {Aug},
  publisher = {American Physical Society},
  doi = {10.1103/PhysRevLett.99.067004},
  url = {https://link.aps.org/doi/10.1103/PhysRevLett.99.067004}
}

@article{PhysRevB.103.245302,
  title = {Theory of the nonreciprocal {J}osephson effect},
  author = {Misaki, Kou and Nagaosa, Naoto},
  journal = {Phys. Rev. B},
  volume = {103},
  issue = {24},
  pages = {245302},
  numpages = {10},
  year = {2021},
  month = {Jun},
  publisher = {American Physical Society},
  doi = {10.1103/PhysRevB.103.245302},
  url = {https://link.aps.org/doi/10.1103/PhysRevB.103.245302}
}

@article{PhysRevB.106.214524,
  title = {Theory of giant diode effect in $d$-wave superconductor junctions on the surface of a topological insulator},
  author = {Tanaka, Yukio and Lu, Bo and Nagaosa, Naoto},
  journal = {Phys. Rev. B},
  volume = {106},
  issue = {21},
  pages = {214524},
  numpages = {13},
  year = {2022},
  month = {Dec},
  publisher = {American Physical Society},
  doi = {10.1103/PhysRevB.106.214524},
  url = {https://link.aps.org/doi/10.1103/PhysRevB.106.214524}
}

@article{PhysRevB.111.184515,
  title = {Orientation-dependent transport in junctions formed by $d$-wave altermagnets and $d$-wave superconductors},
  author = {Zhao, Wenjun and Fukaya, Yuri and Burset, Pablo and Cayao, Jorge and Tanaka, Yukio and Lu, Bo},
  journal = {Phys. Rev. B},
  volume = {111},
  issue = {18},
  pages = {184515},
  numpages = {11},
  year = {2025},
  month = {May},
  publisher = {American Physical Society},
  doi = {10.1103/PhysRevB.111.184515},
  url = {https://link.aps.org/doi/10.1103/PhysRevB.111.184515}
}

@article{PhysRevB.110.014519,
  title = {{J}osephson diode effect in topological superconductors},
  author = {Liu, Zhaochen and Huang, Linghao and Wang, Jing},
  journal = {Phys. Rev. B},
  volume = {110},
  issue = {1},
  pages = {014519},
  numpages = {7},
  year = {2024},
  month = {Jul},
  publisher = {American Physical Society},
  doi = {10.1103/PhysRevB.110.014519},
  url = {https://link.aps.org/doi/10.1103/PhysRevB.110.014519}
}

@article{PhysRevLett.133.226002,
  title = {$\ensuremath{\varphi}$ {J}osephson Junction Induced by Altermagnetism},
  author = {Lu, Bo and Maeda, Kazuki and Ito, Hiroyuki and Yada, Keiji and Tanaka, Yukio},
  journal = {Phys. Rev. Lett.},
  volume = {133},
  issue = {22},
  pages = {226002},
  numpages = {6},
  year = {2024},
  month = {Nov},
  publisher = {American Physical Society},
  doi = {10.1103/PhysRevLett.133.226002},
  url = {https://link.aps.org/doi/10.1103/PhysRevLett.133.226002}
}

@article{PhysRevB.111.064502,
  title = {{J}osephson effect and odd-frequency pairing in superconducting junctions with unconventional magnets},
  author = {Fukaya, Yuri and Maeda, Kazuki and Yada, Keiji and Cayao, Jorge and Tanaka, Yukio and Lu, Bo},
  journal = {Phys. Rev. B},
  volume = {111},
  issue = {6},
  pages = {064502},
  numpages = {15},
  year = {2025},
  month = {Feb},
  publisher = {American Physical Society},
  doi = {10.1103/PhysRevB.111.064502},
  url = {https://link.aps.org/doi/10.1103/PhysRevB.111.064502}
}

@article{PhysRevResearch.5.033199,
  title = {{J}osephson diode effect in {A}ndreev molecules},
  author = {Pillet, J.-D. and Annabi, S. and Peugeot, A. and Riechert, H. and Arrighi, E. and Griesmar, J. and Bretheau, L.},
  journal = {Phys. Rev. Res.},
  volume = {5},
  issue = {3},
  pages = {033199},
  numpages = {6},
  year = {2023},
  month = {Sep},
  publisher = {American Physical Society},
  doi = {10.1103/PhysRevResearch.5.033199},
  url = {https://link.aps.org/doi/10.1103/PhysRevResearch.5.033199}
}

@article{PhysRevB.110.235426,
  title = {Non-{H}ermitian multiterminal phase-biased {J}osephson junctions},
  author = {Cayao, Jorge and Sato, Masatoshi},
  journal = {Phys. Rev. B},
  volume = {110},
  issue = {23},
  pages = {235426},
  numpages = {8},
  year = {2024},
  month = {Dec},
  publisher = {American Physical Society},
  doi = {10.1103/PhysRevB.110.235426},
  url = {https://link.aps.org/doi/10.1103/PhysRevB.110.235426}
}

@Article{10.21468/SciPostPhysCore.2.2.009,
	title={{Scattering description of {A}ndreev molecules}},
	author={Jean-Damien Pillet and Vincent Benzoni and Jo\"{e}l Griesmar and Jean-Loup Smirr and \c{C}a\v{g}lar \"{O}. Girit},
	journal={SciPost Phys. Core},
	volume={2},
	pages={009},
	year={2020},
	publisher={SciPost},
	url={https://scipost.org/10.21468/SciPostPhysCore.2.2.009},
}

@article{PRXQuantum.5.020301,
  title = {Charge Sensing the Parity of an {A}ndreev Molecule},
  author = {van Driel, David and Roovers, Bart and Zatelli, Francesco and Bordin, Alberto and Wang, Guanzhong and van Loo, Nick and Wolff, Jan Cornelis and Mazur, Grzegorz P. and Gazibegovic, Sasa and Badawy, Ghada and Bakkers, Erik P.A.M. and Kouwenhoven, Leo P. and Dvir, Tom},
  journal = {PRX Quantum},
  volume = {5},
  issue = {2},
  pages = {020301},
  numpages = {14},
  year = {2024},
  month = {Apr},
  publisher = {American Physical Society},
  doi = {10.1103/PRXQuantum.5.020301},
  url = {https://link.aps.org/doi/10.1103/PRXQuantum.5.020301}
}

@article{PhysRevB.108.174502,
  title = {On-off switch and sign change for a nonlocal {J}osephson diode in spin-valve {A}ndreev molecules},
  author = {Hodt, Erik Wegner and Linder, Jacob},
  journal = {Phys. Rev. B},
  volume = {108},
  issue = {17},
  pages = {174502},
  numpages = {8},
  year = {2023},
  month = {Nov},
  publisher = {American Physical Society},
  doi = {10.1103/PhysRevB.108.174502},
  url = {https://link.aps.org/doi/10.1103/PhysRevB.108.174502}
}

@article{Matsuo_2022,
   title={Observation of nonlocal {J}osephson effect on double {InAs} nanowires},
   volume={5},
   ISSN={2399-3650},
   url={http://dx.doi.org/10.1038/s42005-022-00994-0},
    pages={221},
   journal={Commun. Phys.},
    author={Matsuo, Sadashige and Lee, Joon Sue and Chang, Chien-Yuan and Sato, Yosuke and Ueda, Kento and Palmstr{\o}m, Christopher J. and Tarucha, Seigo},
   year={2022} }

@article{matsuo2023phase,
  title={Phase engineering of anomalous {J}osephson effect derived from {A}ndreev molecules},
  author={Matsuo, Sadashige and Imoto, Takaya and Yokoyama, Tomohiro and Sato, Yosuke and Lindemann, Tyler and Gronin, Sergei and Gardner, Geoffrey C and Manfra, Michael J and Tarucha, Seigo},
  journal={Sci. Adv.},
  volume={9},
  number={50},
  pages={eadj3698},
  year={2023},
  url={https://doi.org/10.1126/sciadv.adj3698}
}

@article{kotetes2024nonRecifourpi,
	title = {Nonreciprocal equilibrium {J}osephson effect of arbitrary periodicity from {p}oor {m}an's {M}ajorana zero modes},
	author = {Kotetes, Panagiotis and Roig, Merc\`e and Andersen, Brian M.},
	journal = {Phys. Rev. B},
	volume = {113},
	issue = {24},
	pages = {L241403},
	numpages = {8},
	year = {2026},
	month = {Jun},
	publisher = {American Physical Society},
	url = {https://link.aps.org/doi/10.1103/jt2g-hbwq}
}

@article{clarke2008SC,
  title={Superconducting quantum bits},
  author={Clarke, John and Wilhelm, Frank K},
  journal={Nature},
  volume={453},
  number={7198},
  pages={1031--1042},
  year={2008},
  url={https://www.nature.com/articles/nature07128}
}

@article{Brecht2016,
  title={Multilayer microwave integrated quantum circuits for scalable quantum computing},
  author={Brecht, Teresa and Pfaff, Wolfgang and Wang, Chen and Chu, Yiwen and Frunzio, Luigi and Devoret, Michel H and Schoelkopf, Robert J},
  journal={npj Quantum Inf.},
  volume={2},
  number={1},
  pages={16002},
  year={2016},
  url={https://doi.org/10.1038/npjqi.2016.2}
}

@article{fukaya2026crossed,
	title={Crossed surface flat bands in three-dimensional superconducting altermagnets},
	author={Fukaya, Yuri and Lu, Bo and Yada, Keiji and Tanaka, Yukio and Cayao, Jorge},
	journal={Phys. Rev. Lett.},
	volume={136},
	number={22},
	pages={226001},
	year={2026},
	publisher={APS},
	url={https://doi.org/10.1103/65q6-5wxl}
}

@article{Su_2017,
   title={{A}ndreev molecules in semiconductor nanowire double quantum dots},
   volume={8},
    url={http://dx.doi.org/10.1038/s41467-017-00665-7},
    pages={585},
   journal={Nat. Commun.},
   author={Su, Zhaoen and Tacla, Alexandre B. and Hocevar, Mo\"{i}ra and Car, Diana and Plissard, S\'{e}bastien R. and Bakkers, Erik P. A. M. and Daley, Andrew J. and Pekker, David and Frolov, Sergey M.},
   year={2017} 
    }

@article{strambini2016omega,
  title={The $\omega$-SQUIPT as a tool to phase-engineer {J}osephson topological materials},
  author={Strambini, E and D'ambrosio, S and Vischi, F and Bergeret, FS and Nazarov, Yu V and Giazotto, F},
  journal={Nat. Nanotech.},
  volume={11},
  number={12},
  pages={1055--1059},
  year={2016},
  publisher={Nature Publishing Group UK London},
  url={https://doi.org/10.1038/nnano.2016.157}
}

@article{graziano2022selective,
  title={Selective control of conductance modes in multi-terminal {J}osephson junctions},
  author={Graziano, Gino V and Gupta, Mohit and Pendharkar, Mihir and Dong, Jason T and Dempsey, Connor P and Palmstr{\o}m, Chris and Pribiag, Vlad S},
  journal={Nat. Commun.},
  volume={13},
  number={1},
  pages={5933},
  year={2022},
  url={https://doi.org/10.1038/s41467-022-33682-2}
}

@article{martinis2020quantum,
  title={Quantum {J}osephson junction circuits and the dawn of artificial atoms},
  author={Martinis, John M and Devoret, Michel H and Clarke, John},
  journal={Nat. Phys.},
  volume={16},
  number={3},
  pages={234--237},
  year={2020},
  url={https://www.nature.com/articles/s41567-020-0829-5}
}

@article{draelos2019supercurrent,
  title={Supercurrent flow in multiterminal graphene {J}osephson junctions},
  author={Draelos, Anne W and Wei, Ming-Tso and Seredinski, Andrew and Li, Hengming and Mehta, Yash and Watanabe, Kenji and Taniguchi, Takashi and Borzenets, Ivan V and Amet, Fran{\c{c}}ois and Finkelstein, Gleb},
  journal={Nano Lett.},
  volume={19},
  number={2},
  pages={1039--1043},
  year={2019},
  url={https://pubs.acs.org/doi/10.1021/acs.nanolett.8b04330}
}

@article{PhysRevB.97.035443,
  title = {Weyl nodes in {A}ndreev spectra of multiterminal {J}osephson junctions: {C}hern numbers, conductances, and supercurrents},
  author = {Xie, Hong-Yi and Vavilov, Maxim G. and Levchenko, Alex},
  journal = {Phys. Rev. B},
  volume = {97},
  issue = {3},
  pages = {035443},
  numpages = {8},
  year = {2018},
  month = {Jan},
  publisher = {American Physical Society},
  doi = {10.1103/PhysRevB.97.035443},
  url = {https://link.aps.org/doi/10.1103/PhysRevB.97.035443}
}

@article{Haxell_2023,
   title={Demonstration of the Nonlocal {J}osephson Effect in {A}ndreev Molecules},
    author={Haxell, Daniel Z. and Coraiola, Marco and Hinderling, Manuel and ten Kate, Sofieke C. and Sabonis, Deividas and Svetogorov, Aleksandr E. and Belzig, Wolfgang and Cheah, Erik and Krizek, Filip and Schott, R\"{u}diger and Wegscheider, Werner and Nichele, Fabrizio},
   volume={23},
      number={16},
      pages={7532},
   url={http://dx.doi.org/10.1021/acs.nanolett.3c02066},
   journal={Nano Lett.},
   publisher={American Chemical Society (ACS)},
   year={2023}
}

@article{PRXQuantum.5.020340,
  title = {Quantum Circuits with Multiterminal {J}osephson-{A}ndreev Junctions},
  author = {Matute-Ca\~nadas, F.J. and Tosi, L. and Yeyati, A. Levy},
  journal = {PRX Quantum},
  volume = {5},
  issue = {2},
  pages = {020340},
  numpages = {25},
  year = {2024},
  month = {May},
  publisher = {American Physical Society},
  doi = {10.1103/PRXQuantum.5.020340},
  url = {https://link.aps.org/doi/10.1103/PRXQuantum.5.020340}
}

@Article{10.21468/SciPostPhys.17.2.037,
	title={{Theory of universal diode effect in three-terminal {J}osephson junctions}},
	author={Jorge Huamani Correa and Michal P. Nowak},
	journal={SciPost Phys.},
	volume={17},
	pages={037},
	year={2024},
	publisher={SciPost},
	doi={10.21468/SciPostPhys.17.2.037},
	url={https://scipost.org/10.21468/SciPostPhys.17.2.037}
	}

@article{riwar2016multi,
  title={Multi-terminal {J}osephson junctions as topological matter},
  author={Riwar, Roman-Pascal and Houzet, Manuel and Meyer, Julia S and Nazarov, Yuli V},
  journal={Nat. Commun.},
  volume={7},
  number={1},
  pages={11167},
  year={2016},
  publisher={Nature Publishing Group UK London},
  url={https://doi.org/10.1038/ncomms11167}
}

@article{matsuo2023phase2,
  title={Phase-dependent {A}ndreev molecules and superconducting gap closing in coherently-coupled {J}osephson junctions},
  author={Matsuo, Sadashige and Imoto, Takaya and Yokoyama, Tomohiro and Sato, Yosuke and Lindemann, Tyler and Gronin, Sergei and Gardner, Geoffrey C and Nakosai, Sho and Tanaka, Yukio and Manfra, Michael J and others},
  journal={Nat. Commun.},
  volume={14},
  number={1},
  pages={8271},
  year={2023},
  publisher={Nature Publishing Group UK London},
  url={https://doi.org/10.1038/s41467-023-44111-3}
}

@article{tanakaReview2024,
    author = {Tanaka, Yukio and Tamura, Shun and Cayao, Jorge},
    title = {Theory of {M}ajorana Zero Modes in Unconventional Superconductors},
    journal = {Prog. Theor. Exp. Phys.},
    volume={2024},
    pages = {08C105},
    year = {2024},
    month = {05},
    doi = {10.1093/ptep/ptae065},
    url = {https://doi.org/10.1093/ptep/ptae065}
}

@article{FukayaJPCM2025,
    author = {Fukaya, Yuri and Lu, Bo and Yada, Keiji and Tanaka, Yukio and Cayao, Jorge},
    title = {Superconducting phenomena in systems with unconventional magnets},
    journal = {J. Phys.: Condens. Matter},
    volume = {37},
    pages = {313003},
        year ={2025},
        url={https://iopscience.iop.org/article/10.1088/1361-648X/adf1cf}
}

@article{Pillet_2019,
   title={Nonlocal {J}osephson Effect in {A}ndreev Molecules},
   volume={19},
   url={http://dx.doi.org/10.1021/acs.nanolett.9b02686},
   number={10},
   journal={Nano Lett.},
   publisher={American Chemical Society (ACS)},
   author={Pillet, J.-D. and Benzoni, V. and Griesmar, J. and Smirr, J.-L. and Girit, \c{C}. \"{O}.},
pages={7138},
   year={2019} }

@article{PhysRevX.10.031051,
  title = {Multiterminal {J}osephson Effect},
  author = {Pankratova, Natalia and Lee, Hanho and Kuzmin, Roman and Wickramasinghe, Kaushini and Mayer, William and Yuan, Joseph and Vavilov, Maxim G. and Shabani, Javad and Manucharyan, Vladimir E.},
  journal = {Phys. Rev. X},
  volume = {10},
  issue = {3},
  pages = {031051},
  numpages = {12},
  year = {2020},
  month = {Sep},
  publisher = {American Physical Society},
  doi = {10.1103/PhysRevX.10.031051},
  url = {https://link.aps.org/doi/10.1103/PhysRevX.10.031051}
}

@article{pita2023direct,
  title={Direct manipulation of a superconducting spin qubit strongly coupled to a transmon qubit},
  author={Pita-Vidal, Marta and Bargerbos, Arno and {\v{Z}}itko, Rok and Splitthoff, Lukas J and Gr{\"u}nhaupt, Lukas and Wesdorp, Jaap J and Liu, Yu and Kouwenhoven, Leo P and Aguado, Ram{\'o}n and van Heck, Bernard and others},
  journal={Nat. Phys.},
    volume={19},
  pages={1110},
  year={2023},
  url={https://doi.org/10.1038/s41567-023-02071-x}
}

@article{siddiqi2021engineering,
  title={Engineering high-coherence superconducting qubits},
  author={Siddiqi, Irfan},
  journal={Nat. Rev. Mater.},
  volume={6},
  number={10},
  pages={875--891},
  year={2021},
  url={https://www.nature.com/articles/s41578-021-00370-4}
}

@article{PRXQuantum.2.040204,
  title = {Superconducting Circuit Companion---an Introduction with Worked Examples},
  author = {Rasmussen, S.E. and Christensen, K.S. and Pedersen, S.P. and Kristensen, L.B. and B\ae{}kkegaard, T. and Loft, N.J.S. and Zinner, N.T.},
  journal = {PRX Quantum},
  volume = {2},
  issue = {4},
  pages = {040204},
  numpages = {58},
  year = {2021},
  month = {Dec},
  publisher = {American Physical Society},
  doi = {10.1103/PRXQuantum.2.040204},
  url = {https://link.aps.org/doi/10.1103/PRXQuantum.2.040204}
}

@article{krantz2019quantum,
  title={A quantum engineer's guide to superconducting qubits},
  author={Krantz, Philip and Kjaergaard, Morten and Yan, Fei and Orlando, Terry P and Gustavsson, Simon and Oliver, William D},
  journal={ Appl. Phys. Rev.},
  volume={6},
  number={2},
  pages={021318},
  year={2019},
  url={https://doi.org/10.1063/1.5089550}
}

@article{bargerbos2022singlet,
  title = {Singlet-Doublet Transitions of a Quantum Dot {J}osephson Junction Detected in a Transmon Circuit},
  author = {Bargerbos, Arno and Pita-Vidal, Marta and \ifmmode \check{Z}\else \v{Z}\fi{}itko, Rok and \'Avila, Jes\'us and Splitthoff, Lukas J. and Gr\"unhaupt, Lukas and Wesdorp, Jaap J. and Andersen, Christian K. and Liu, Yu and Kouwenhoven, Leo P. and Aguado, Ram\'on and Kou, Angela and van Heck, Bernard},
  journal = {PRX Quantum},
  volume = {3},
  issue = {3},
  pages = {030311},
  numpages = {14},
  year = {2022},
  month = {Jul},
  publisher = {American Physical Society},
  doi = {10.1103/PRXQuantum.3.030311},
  url = {https://link.aps.org/doi/10.1103/PRXQuantum.3.030311}
}

@article{doi:10.1146/annurev-conmatphys-031119-050605,
author = {Kjaergaard, Morten and Schwartz, Mollie E. and Braum\"{u}ller, Jochen and Krantz, Philip and Wang, Joel I.-J. and Gustavsson, Simon and Oliver, William D.},
title = {Superconducting Qubits: Current State of Play},
journal = {Annu. Rev. Condens. Matter Phys.},
volume = {11},
number = {1},
pages = {369-395},
year = {2020},
doi = {10.1146/annurev-conmatphys-031119-050605},
URL = { https://doi.org/10.1146/annurev-conmatphys-031119-050605}
}

@article{PhysRevB.111.024506,
  title = {Gate-tunable nonlocal {J}osephson effect through magnetic van der {W}aals bilayers},
  author = {Bobkov, G. A. and Rabinovich, D. S. and Bobkov, A. M. and Bobkova, I. V.},
  journal = {Phys. Rev. B},
  volume = {111},
  issue = {2},
  pages = {024506},
  numpages = {11},
  year = {2025},
  month = {Jan},
  publisher = {American Physical Society},
  doi = {10.1103/PhysRevB.111.024506},
  url = {https://link.aps.org/doi/10.1103/PhysRevB.111.024506}
}

@article{benito2020hybrid,
  title={Hybrid superconductor-semiconductor systems for quantum technology},
  author={Benito, M\'{o}nica and Burkard, Guido},
  journal={Appl. Phys. Lett.},
  volume={116},
  number={19},
  pages={190502},
  year={2020},
  url={https://doi.org/10.1063/5.0004777}
}

@article{junger2023intermediate,
  title={Intermediate states in {A}ndreev bound state fusion},
  author={J{\"u}nger, Christian and Lehmann, Sebastian and Dick, Kimberly A and Thelander, Claes and Sch{\"o}nenberger, Christian and Baumgartner, Andreas},
  journal={Commun. Phys.},
  volume={6},
  number={1},
  pages={190},
  year={2023},
  publisher={Nature Publishing Group UK London},
  url={https://doi.org/10.1038/s42005-023-01273-2}
}

@misc{devoret2004superconducting,
      title={Superconducting Qubits: A Short Review}, 
      author={M. H. Devoret and A. Wallraff and J. M. Martinis},
      year={2004},
      eprint={cond-mat/0411174},
      archivePrefix={arXiv},
      primaryClass={cond-mat.mes-hall}
}

@article{cayao2026nonlocal,
	title={Nonlocal Josephson diode effect in minimal Kitaev chains},
	author={Cayao, Jorge and Sato, Masatoshi},
	journal={Phys. Rev. Res.},
	volume={8},
	number={1},
	pages={013326},
	year={2026},
	publisher={APS},
	url={https://doi.org/10.1103/hf7s-f7tj}
}

@article{Furusaki_1999,
 doi = {10.1006/spmi.1999.0730},
 url = {https://doi.org/10.1006\%2Fspmi.1999.0730},
 year = 1999,
 month = {may},
 publisher = {Elsevier {BV}},
 volume = {25},
 number = {5-6},
 pages = {809--818},
 author = {Akira Furusaki},
 title = {{J}osephson current carried by {A}ndreev levels in superconducting quantum point contacts},
 journal = {Superlattices and Microstruct.}
}

@conference{Beenakker:92,
	Author = {Beenakker, CWJ},
	Booktitle = {Transport {P}henomena in {M}esoscopic {S}ystems: {P}roceedings of the 14th {T}aniguchi {S}ymposium, Shima, Japan, November 10-14, 1991},
	Publisher = {Springer, Berlin},
	Volume = {109},
	Pages = {235},
	Title = {Three ``Universal" Mesoscopic {J}osephson Effects},
	Year = {1992}}

@article{furusaki1991dc,
  title={Dc {J}osephson effect and {A}ndreev reflection},
  author={Furusaki, Akira and Tsukada, Masaru},
  journal={Solid State Commun.},
  volume={78},
  number={4},
  pages={299--302},
  year={1991},
  publisher={Elsevier},
  url = {https://doi.org/10.1016/0038-1098(91)90201-6}
}

@article{cayao2018andreev,
  title={{A}ndreev spectrum and supercurrents in nanowire-based {SNS} junctions containing {M}ajorana bound states},
  author={Cayao, Jorge and Black-Schaffer, Annica M and Prada, Elsa and Aguado, Ram{\'o}n},
  journal={Beilstein J. Nanotechnol.},
  volume={9},
  number={1},
  pages={1339--1357},
  year={2018},
  publisher={Beilstein-Institut},
  url={https://www.beilstein-journals.org/bjnano/articles/9/127},
}

@article{PhysRevLett.123.117001,
  title = {Supercurrent Detection of Topologically Trivial Zero-Energy States in Nanowire Junctions},
  author = {Awoga, Oladunjoye A. and Cayao, Jorge and Black-Schaffer, Annica M.},
  journal = {Phys. Rev. Lett.},
  volume = {123},
  issue = {11},
  pages = {117001},
  numpages = {7},
  year = {2019},
  month = {Sep},
  publisher = {American Physical Society},
  doi = {10.1103/PhysRevLett.123.117001},
  url = {https://link.aps.org/doi/10.1103/PhysRevLett.123.117001}
}

@article{shaffer2025SDE,
      title={Theories of Superconducting Diode Effects}, 
      author={Daniel Shaffer and Alex Levchenko},
      year={2025},
      journal={arXiv:2510.25864},
      url={https://arxiv.org/abs/2510.25864} 
}

@article{PhysRevB.109.205406,
  title = {Controllable odd-frequency {C}ooper pairs in multisuperconductor {J}osephson junctions},
  author = {Cayao, Jorge and Burset, Pablo and Tanaka, Yukio},
  journal = {Phys. Rev. B},
  volume = {109},
  issue = {20},
  pages = {205406},
  numpages = {10},
  year = {2024},
  month = {May},
  publisher = {American Physical Society},
  doi = {10.1103/PhysRevB.109.205406},
  url = {https://link.aps.org/doi/10.1103/PhysRevB.109.205406}
}

@article{PhysRevX.7.021032,
  title = {Topological Superconductivity in a Planar {J}osephson Junction},
  author = {Pientka, Falko and Keselman, Anna and Berg, Erez and Yacoby, Amir and Stern, Ady and Halperin, Bertrand I.},
  journal = {Phys. Rev. X},
  volume = {7},
  issue = {2},
  pages = {021032},
  numpages = {17},
  year = {2017},
  month = {May},
  publisher = {American Physical Society},
  doi = {10.1103/PhysRevX.7.021032},
  url = {https://link.aps.org/doi/10.1103/PhysRevX.7.021032}
}

@article{PhysRevB.104.075402,
  title = {Ultralong-distance quantum correlations in three-terminal {J}osephson junctions},
  author = {M\'elin, R\'egis},
  journal = {Phys. Rev. B},
  volume = {104},
  issue = {7},
  pages = {075402},
  numpages = {19},
  year = {2021},
  month = {Aug},
  publisher = {American Physical Society},
  doi = {10.1103/PhysRevB.104.075402},
  url = {https://link.aps.org/doi/10.1103/PhysRevB.104.075402}
}

@article{PhysRevB.102.245435,
  title = {Inversion in a four-terminal superconducting device on the quartet line. I. Two-dimensional metal and the quartet beam splitter},
  author = {M\'elin, R\'egis},
  journal = {Phys. Rev. B},
  volume = {102},
  issue = {24},
  pages = {245435},
  numpages = {30},
  year = {2020},
  month = {Dec},
  publisher = {American Physical Society},
  doi = {10.1103/PhysRevB.102.245435},
  url = {https://link.aps.org/doi/10.1103/PhysRevB.102.245435}
}

@article{PhysRevB.96.205425,
  title = {{M}ajorana splitting from critical currents in {J}osephson junctions},
  author = {Cayao, Jorge and San-Jose, Pablo and Black-Schaffer, Annica M. and Aguado, Ram\'on and Prada, Elsa},
  journal = {Phys. Rev. B},
  volume = {96},
  issue = {20},
  pages = {205425},
  numpages = {9},
  year = {2017},
  month = {Nov},
  publisher = {American Physical Society},
  doi = {10.1103/PhysRevB.96.205425},
  url = {https://link.aps.org/doi/10.1103/PhysRevB.96.205425}
}

@article{PhysRevB.104.L020501,
  title = {Distinguishing trivial and topological zero-energy states in long nanowire junctions},
  author = {Cayao, Jorge and Black-Schaffer, Annica M.},
  journal = {Phys. Rev. B},
  volume = {104},
  issue = {2},
  pages = {L020501},
  numpages = {6},
  year = {2021},
  month = {Jul},
  publisher = {American Physical Society},
  doi = {10.1103/PhysRevB.104.L020501},
  url = {https://link.aps.org/doi/10.1103/PhysRevB.104.L020501}
}

@book{zagoskin,
	Author = {Alexandre Zagoskin},
	title = {Quantum {T}heory of {M}any-{B}ody {S}ystems: {T}echniques and {A}pplications},
        publisher={Springer, Berlin},
	Year = {2014}}

@article{RevModPhys.76.411,
  title = {The current-phase relation in {J}osephson junctions},
  author = {Golubov, A. A. and Kupriyanov, M. Yu. and Il'ichev, E.},
  journal = {Rev. Mod. Phys.},
  volume = {76},
  issue = {2},
  pages = {411--469},
  numpages = {0},
  year = {2004},
  month = {Apr},
  publisher = {American Physical Society},
  doi = {10.1103/RevModPhys.76.411},
  url = {https://link.aps.org/doi/10.1103/RevModPhys.76.411}
}

@article{noda2016momentum,
  title={Momentum-dependent band spin splitting in semiconducting MnO 2: a density functional calculation},
  author={Noda, Yusuke and Ohno, Kaoru and Nakamura, Shinichiro},
  journal={Phys. Chem. Chem. Phys.},
  volume={18},
  number={19},
  pages={13294--13303},
  year={2016},
  url={https://doi.org/10.1039/C5CP07806G},
  publisher={Royal Society of Chemistry}
}

@article{NakaNatCommun2019,
    author = {Naka, Makoto and  Hayami, Satoru and Kusunose, Hiroaki and Yanagi, Yuki and Motome, Yukitoshi and Seo, Hitoshi},
    title = {Spin current generation in organic antiferromagnets},
    journal = {Nat. Commun.},
    year = {2019},
    volume = {10},
    pages = {4305},
    url = {https://doi.org/10.1038/s41467-019-12229-y}
}

@article{NakaPRB2020,
  title = {Anomalous {H}all effect in $\ensuremath{\kappa}$-type organic antiferromagnets},
  author = {Naka, Makoto and Hayami, Satoru and Kusunose, Hiroaki and Yanagi, Yuki and Motome, Yukitoshi and Seo, Hitoshi},
  journal = {Phys. Rev. B},
  volume = {102},
  issue = {7},
  pages = {075112},
  numpages = {11},
  year = {2020},
  month = {Aug},
  publisher = {American Physical Society},
  doi = {10.1103/PhysRevB.102.075112},
  url = {https://link.aps.org/doi/10.1103/PhysRevB.102.075112}
}

@article{Yuanprm21,
  title = {Prediction of low-$Z$ collinear and noncollinear antiferromagnetic compounds having momentum-dependent spin splitting even without spin-orbit coupling},
  author = {Yuan, Lin-Ding and Wang, Zhi and Luo, Jun-Wei and Zunger, Alex},
  journal = {Phys. Rev. Mater.},
  volume = {5},
  issue = {1},
  pages = {014409},
  numpages = {24},
  year = {2021},
  month = {Jan},
  publisher = {American Physical Society},
  doi = {10.1103/PhysRevMaterials.5.014409},
  url = {https://link.aps.org/doi/10.1103/PhysRevMaterials.5.014409}
}

@article{Hayami19,
	author = {Hayami ,Satoru and Yanagi ,Yuki and Kusunose ,Hiroaki},
	title = {Momentum-Dependent Spin Splitting by Collinear Antiferromagnetic Ordering},
	journal = {J. Phys. Soc. Jpn.},
	volume = {88},
	number = {12},
	pages = {123702},
	year = {2019},
	doi = {10.7566/JPSJ.88.123702},
	URL = { https://doi.org/10.7566/JPSJ.88.123702}
}

@article{kashiwaya2000,
	title={Tunnelling effects on surface bound states in unconventional superconductors},
	author={Kashiwaya, Satoshi and Tanaka, Yukio},
	journal={Rep. Prog. Phys.},
	volume={63},
	number={10},
	pages={1641},
	year={2000},
	publisher={IOP Publishing},
	url={https://doi.org/10.1088/0034-4885/63/10/202},
}

@article{aguado2020perspective,
	title={A perspective on semiconductor-based superconducting qubits},
	author={Aguado, Ram{\'o}n},
	journal={Appl. Phys. Lett. },
	volume={117},
	number={24},
	pages={240501},
	year={2020},
	url={https://doi.org/10.1063/5.0024124},
	publisher={AIP Publishing}
}

@article{aguado2020majorana,
	title={{M}ajorana qubits for topological quantum computing},
	author={Aguado, Ram{\'o}n and Kouwenhoven, Leo P},
	journal={Phys. Today},
	volume={73},
	number={6},
	pages={44--50},
	year={2020},
	url={https://doi.org/10.1063/PT.3.4499},
	publisher={American Institute of Physics}
}

@article{PhysRevB.92.035428,
	title = {Topological {J}osephson ${\ensuremath{\phi}}_{0}$ junctions},
	author = {Dolcini, Fabrizio and Houzet, Manuel and Meyer, Julia S.},
	journal = {Phys. Rev. B},
	volume = {92},
	issue = {3},
	pages = {035428},
	numpages = {7},
	year = {2015},
	month = {Jul},
	publisher = {American Physical Society},
	doi = {10.1103/PhysRevB.92.035428},
	url = {https://link.aps.org/doi/10.1103/PhysRevB.92.035428}
}

@article{PhysRevB.93.174502,
	title = {Anomalous {J}osephson effect in semiconducting nanowires as a signature of the topologically nontrivial phase},
	author = {Nesterov, Konstantin N. and Houzet, Manuel and Meyer, Julia S.},
	journal = {Phys. Rev. B},
	volume = {93},
	issue = {17},
	pages = {174502},
	numpages = {9},
	year = {2016},
	month = {May},
	publisher = {American Physical Society},
	doi = {10.1103/PhysRevB.93.174502},
	url = {https://link.aps.org/doi/10.1103/PhysRevB.93.174502}
}

@article{79tj-c3y4,
	title = {{J}osephson diode effect with {A}ndreev and {M}ajorana bound states},
	author = {Mondal, Sayan and Fu, Pei-Hao and Cayao, Jorge},
	journal = {Phys. Rev. B},
	volume = {112},
	issue = {14},
	pages = {144506},
	numpages = {16},
	year = {2025},
	month = {Oct},
	publisher = {American Physical Society},
	doi = {10.1103/79tj-c3y4},
	url = {https://link.aps.org/doi/10.1103/79tj-c3y4}
}

@article{zhu2025josephson,
	title={{J}osephson diode effect in nanowire-based {A}ndreev molecules},
	author={Zhu, Shang and Ma, Yiwen and He, Jiangbo and Yang, Xiaozhou and Jia, Zhongmou and Wei, Min and Jiao, Yiping and He, Jiezhong and Zhuo, Enna and Cao, Xuewei and others},
	journal={Commun. Phys.},
	volume={8},
	number={1},
	pages={330},
	year={2025},
	publisher={Nature Publishing Group UK London},
	url={https://doi.org/10.1038/s42005-025-02237-4}
}

@article{PhysRevB.108.075425,
	title = {Phase-shifted {A}ndreev levels in an altermagnet {J}osephson junction},
	author = {Beenakker, C. W. J. and Vakhtel, T.},
	journal = {Phys. Rev. B},
	volume = {108},
	issue = {7},
	pages = {075425},
	numpages = {7},
	year = {2023},
	month = {Aug},
	publisher = {American Physical Society},
	doi = {10.1103/PhysRevB.108.075425},
	url = {https://link.aps.org/doi/10.1103/PhysRevB.108.075425}
}

@article{LiborPRX22,
	title = {Beyond Conventional Ferromagnetism and Antiferromagnetism: A Phase with Nonrelativistic Spin and Crystal Rotation Symmetry},
	author = {\v{S}mejkal, Libor and Sinova, Jairo and Jungwirth, Tomas},
	journal = {Phys. Rev. X},
	volume = {12},
	issue = {3},
	pages = {031042},
	numpages = {16},
	year = {2022},
	publisher = {American Physical Society},
	doi = {10.1103/PhysRevX.12.031042},
	url = {https://link.aps.org/doi/10.1103/PhysRevX.12.031042}
}

@article{Yuanprb20,
  title = {Giant momentum-dependent spin splitting in centrosymmetric low-$Z$ antiferromagnets},
  author = {Yuan, Lin-Ding and Wang, Zhi and Luo, Jun-Wei and Rashba, Emmanuel I. and Zunger, Alex},
  journal = {Phys. Rev. B},
  volume = {102},
  issue = {1},
  pages = {014422},
  numpages = {13},
  doi = {10.1103/PhysRevB.102.014422},
  url = {https://link.aps.org/doi/10.1103/PhysRevB.102.014422},
    year = {2020}
}

@article{Ahn2019,
	title = {Antiferromagnetism in ${\mathrm{RuO}}_{2}$ as $d$-wave {P}omeranchuk instability},
	author = {Ahn, Kyo-Hoon and Hariki, Atsushi and Lee, Kwan-Woo and Kune\v{s}, Jan},
	journal = {Phys. Rev. B},
	volume = {99},
	issue = {18},
	pages = {184432},
	numpages = {5},
	doi = {10.1103/PhysRevB.99.184432},
	url = {https://link.aps.org/doi/10.1103/PhysRevB.99.184432},
		year = {2019}
}

@article{mazin2025notes,
  title={Notes on altermagnetism and superconductivity},
  author={Mazin, Igor I},
  journal = {AAPPS Bull.},
  volume  = {35},
  pages   = {18},
  year    = {2025},
  doi     = {10.1007/s43673-025-00158-6},
  url     = {https://doi.org/10.1007/s43673-025-00158-6}
}

@article{liu2025review,
      title={Altermagnetism and Superconductivity: A Short Historical Review}, 
      author={Zhao Liu and Hui Hu and Xia-Ji Liu},
      year={2025},
      pages={2510.09170},
      journal={arXiv},
      url={https://arxiv.org/abs/2510.09170} 
}

@article{landscape22,
	title = {Emerging Research Landscape of Altermagnetism},
	author = {\v{S}mejkal, Libor and Sinova, Jairo and Jungwirth, Tomas},
	journal = {Phys. Rev. X},
	volume = {12},
	issue = {4},
	pages = {040501},
	numpages = {27},
	year = {2022},
	month = {Dec},
	publisher = {American Physical Society},
	url = {https://link.aps.org/doi/10.1103/PhysRevX.12.040501}
}

@article{LiborSAv,
	author = {Libor \v{S}mejkal  and Rafael Gonz\'{a}lez-Hern\'{a}ndez  and T. Jungwirth  and J. Sinova },
	title = {Crystal time-reversal symmetry breaking and spontaneous {H}all effect in collinear antiferromagnets},
	journal = {Sci. Adv.},
	volume = {6},
	number = {23},
	pages = {eaaz8809},
	year = {2020},
	doi = {10.1126/sciadv.aaz8809},
	URL = {https://www.science.org/doi/abs/10.1126/sciadv.aaz8809},
}

@article{PhysRevLett.131.076003,
	title = {dc {J}osephson Effect in Altermagnets},
	author = {Ouassou, Jabir Ali and Brataas, Arne and Linder, Jacob},
	journal = {Phys. Rev. Lett.},
	volume = {131},
	issue = {7},
	pages = {076003},
	numpages = {6},
	year = {2023},
	month = {Aug},
	publisher = {American Physical Society},
	url = {https://link.aps.org/doi/10.1103/PhysRevLett.131.076003}
}

@article{MazinPRX22,
	title = {Editorial: Altermagnetism---A New Punch Line of Fundamental Magnetism},
	author = {Mazin, Igor},
	collaboration = {The PRX Editors},
	journal = {Phys. Rev. X},
	volume = {12},
	issue = {4},
	pages = {040002},
	numpages = {3},
	year = {2022},
	month = {Dec},
	publisher = {American Physical Society},
	url = {https://link.aps.org/doi/10.1103/PhysRevX.12.040002}
}

@Article{fu2025light,
	title={{Light-induced {F}loquet spin-triplet Cooper pairs in unconventional magnets}},
	author={Pei-Hao Fu and Sayan Mondal and Jun-Feng Liu and Jorge Cayao},
	journal={SciPost Phys.},
	volume={20},
	pages={059},
	year={2026},
	publisher={SciPost},
	doi={10.21468/SciPostPhys.20.2.059},
	url={https://scipost.org/10.21468/SciPostPhys.20.2.059},
}

@article{fu2025floquet,
	title = {Floquet Engineering Spin Triplet States in Unconventional Magnets},
	author = {Fu, Pei-Hao and Mondal, Sayan and Liu, Jun-Feng and Tanaka, Yukio and Cayao, Jorge},
	journal = {Phys. Rev. Lett.},
	volume = {136},
	issue = {6},
	pages = {066703},
	numpages = {11},
	year = {2026},
	month = {Feb},
	publisher = {American Physical Society},
	doi = {10.1103/lkf9-jgv6},
	url = {https://link.aps.org/doi/10.1103/lkf9-jgv6}
}

@article{zhang2024finite,
	title={Finite-momentum {C}ooper pairing in proximitized altermagnets},
	author={Zhang, Song-Bo and Hu, Lun-Hui and Neupert, Titus},
	journal={Nat. Commun.},
	volume={15},
	number={1},
	pages={1801},
	year={2024},
	publisher={Nature Publishing Group UK London},
	url={https://doi.org/10.1038/s41467-024-45951-3}
}

@article{PhysRevB.109.024517,
	title = {Orientation-dependent {J}osephson effect in spin-singlet superconductor/altermagnet/spin-triplet superconductor junctions},
	author = {Cheng, Qiang and Sun, Qing-Feng},
	journal = {Phys. Rev. B},
	volume = {109},
	issue = {2},
	pages = {024517},
	numpages = {10},
	year = {2024},
	month = {Jan},
	publisher = {American Physical Society},
	url = {https://link.aps.org/doi/10.1103/PhysRevB.109.024517}
}

@article{PhysRevB.111.165406,
	title = {Tunable second harmonic in altermagnetic {J}osephson junctions},
	author = {Sun, Hai-Peng and Zhang, Song-Bo and Li, Chang-An and Trauzettel, Bj\"orn},
	journal = {Phys. Rev. B},
	volume = {111},
	issue = {16},
	pages = {165406},
	numpages = {8},
	year = {2025},
	month = {Apr},
	publisher = {American Physical Society},
	url = {https://link.aps.org/doi/10.1103/PhysRevB.111.165406}
}

@article{mj4b-2fnr,
	title = {{A}ndreev bound states and supercurrent in an unconventional superconductor-altermagnet {J}osephson junction},
	author = {Alipourzadeh, Mohammad and Hajati, Yaser},
	journal = {Phys. Rev. B},
	volume = {111},
	issue = {21},
	pages = {214515},
	numpages = {12},
	year = {2025},
	month = {Jun},
	publisher = {American Physical Society},
	url = {https://link.aps.org/doi/10.1103/mj4b-2fnr}
}

@article{Jiang_2025,
	url = {https://doi.org/10.1088/1674-1056/add7aa},
	year = {2025},
	month = {oct},
	publisher = {Chinese Physical Society and IOP Publishing Ltd},
	volume = {34},
	number = {10},
	pages = {107803},
	author = {Jiang, Yi and Liu, Han-Lin and Wang, Jun},
	title = {{J}osephson diode effect in altermagnet-based $s$-wave superconductor junction},
	journal = {Chin. Phys. B},
}

@article{PhysRevB.110.014518,
	title = {Field-free {J}osephson diode effect in altermagnet/normal metal/altermagnet junctions},
	author = {Cheng, Qiang and Mao, Yue and Sun, Qing-Feng},
	journal = {Phys. Rev. B},
	volume = {110},
	issue = {1},
	pages = {014518},
	numpages = {12},
	year = {2024},
	month = {Jul},
	publisher = {American Physical Society},
	url = {https://link.aps.org/doi/10.1103/PhysRevB.110.014518}
}

@article{prnx-47mk,
  title = {{J}osephson current signature of {F}loquet {M}ajorana and topological accidental zero modes in altermagnet heterostructures},
  author = {Pal, Amartya and Mondal, Debashish and Nag, Tanay and Saha, Arijit},
  journal = {Phys. Rev. B},
  volume = {112},
  issue = {20},
  pages = {L201408},
  numpages = {9},
  year = {2025},
  month = {Nov},
  publisher = {American Physical Society},
  doi = {10.1103/prnx-47mk},
  url = {https://link.aps.org/doi/10.1103/prnx-47mk}
}

@article{khodas2026pUM,
	title = {Nonrelativistic {I}sing superconductivity in $p$-wave magnets},
	author = {Khodas, M. and \ifmmode \check{S}\else \v{S}\fi{}mejkal, Libor and Mazin, I. I.},
	journal = {Phys. Rev. B},
	volume = {114},
	issue = {8},
	pages = {L080505},
	numpages = {7},
	year = {2026},
	month = {Aug},
	publisher = {American Physical Society},
	doi = {10.1103/p21c-rtv7},
	url = {https://link.aps.org/doi/10.1103/p21c-rtv7}
}

@article{vosoughinia2025altermon,
      title={Altermon: a magnetic-field-free parity protected qubit based on a narrow altermagnet {J}osephson junction}, 
      author={Sakineh Vosoughi-nia and Michal P. Nowak},
      year={2025},
      journal={arXiv:2510.18145},
      url={https://arxiv.org/abs/2510.18145}
}

@article{esin2026JDE,
	title={{J}osephson diode and spin-valve effects on the surface of altermagnet {C}r{S}b},
	author={Esin, Varnava Denisovich and Kazmin, D Yu and Barash, Yu S and Timonina, Anna Vladimirovna and Kolesnikov, Nikolay Nikolaevich and Deviatov, Eduard Valentinovich},
	journal={JETP Letters},
	volume={123},
	number={8},
	pages={556--565},
	year={2026},
	publisher={Springer},
	url={https://doi.org/10.1134/S0021364026600667}
}

@article{debnath2025,
	title = {Spin polarization and diode effect in thermoelectric current through altermagnet-based superconductor heterostructures},
	author = {Debnath, Debika and Saha, Arijit and Dutta, Paramita},
	journal = {Phys. Rev. B},
	volume = {113},
	issue = {10},
	pages = {104508},
	numpages = {17},
	year = {2026},
	month = {Mar},
	publisher = {American Physical Society},
	doi = {10.1103/jd41-26xr},
	url = {https://link.aps.org/doi/10.1103/jd41-26xr}
}

@article{sharma2026pUM,
      title={$p$-wave magnet driven field-free {J}osephson diode effect}, 
      author={Lovy Sharma and Bimal Ghimire and Manisha Thakurathi},
      year={2026},
      journal={arXiv:2602.16677},
      url={https://arxiv.org/abs/2602.16677} 
}

@article{yqsg-xdg8,
	title = {Tunable {J}osephson diode effect in singlet superconductor-altermagnet-triplet superconductor junctions},
	author = {Sharma, Lovy and Thakurathi, Manisha},
	journal = {Phys. Rev. B},
	volume = {112},
	issue = {10},
	pages = {104506},
	numpages = {10},
	year = {2025},
	month = {Sep},
	publisher = {American Physical Society},
	url = {https://link.aps.org/doi/10.1103/yqsg-xdg8}
}

@article{Maeda2025,
  title = {Classification of pair symmetries in superconductors with unconventional magnetism},
  author = {Maeda, Kazuki and Fukaya, Yuri and Yada, Keiji and Lu, Bo and Tanaka, Yukio and Cayao, Jorge},
  journal = {Phys. Rev. B},
  volume = {111},
  issue = {14},
  pages = {144508},
  numpages = {14},
  year = {2025},
  month = {Apr},
  publisher = {American Physical Society},
  doi = {10.1103/PhysRevB.111.144508},
  url = {https://link.aps.org/doi/10.1103/PhysRevB.111.144508}
}

@article{chakraborty2024,
  title = {Constraints on superconducting pairing in altermagnets},
  author = {Chakraborty, Debmalya and Black-Schaffer, Annica M.},
  journal = {Phys. Rev. B},
  volume = {112},
  issue = {1},
  pages = {014516},
  numpages = {12},
  year = {2025},
  month = {Jul},
  publisher = {American Physical Society},
  doi = {10.1103/zylh-rqxl},
  url = {https://link.aps.org/doi/10.1103/zylh-rqxl}
}

@article{PhysRevB.111.054520,
  title = {Superconducting order parameters in spin space groups: Methodology and application},
  author = {Feng, Xilin and Zhang, Zhongyi},
  journal = {Phys. Rev. B},
  volume = {111},
  issue = {5},
  pages = {054520},
  numpages = {24},
  year = {2025},
  month = {Feb},
  publisher = {American Physical Society},
  doi = {10.1103/PhysRevB.111.054520},
  url = {https://link.aps.org/doi/10.1103/PhysRevB.111.054520}
}

@article{Yokoyama25floquet,
  title = {Floquet engineering triplet superconductivity in superconductors with spin-orbit coupling or altermagnetism},
  author = {Yokoyama, Takehito},
  journal = {Phys. Rev. B},
  volume = {112},
  issue = {2},
  pages = {024512},
  numpages = {10},
  year = {2025},
  month = {Jul},
  publisher = {American Physical Society},
  doi = {10.1103/4tng-rhc4},
  url = {https://link.aps.org/doi/10.1103/4tng-rhc4}
}

\end{document}